\documentclass[10pt,journal,compsoc]{IEEEtran}
\usepackage{paralist}
\usepackage{algorithm}
\usepackage{algorithmic}

\usepackage{amsmath}
\usepackage{multirow}

\usepackage{subfigure}
\usepackage{graphicx}
\usepackage{epsfig}
\usepackage{epstopdf}

\usepackage{amssymb}

\usepackage{bm}
\usepackage{booktabs}
\usepackage{verbatim}
\usepackage{array}
\usepackage{diagbox}

\usepackage[T1]{fontenc}

\ifCLASSOPTIONcompsoc
  \usepackage[nocompress]{cite}
\else
  \usepackage{cite}
\fi

\begin{document}
%
\title{Dual Adversarial Variational Embedding for Robust Recommendation}

\author{Qiaomin~Yi,~
        Ning~Yang,~\IEEEmembership{Member,~IEEE,}
        Philip S.~Yu,~\IEEEmembership{Fellow,~IEEE}
\IEEEcompsocitemizethanks{
\IEEEcompsocthanksitem Qiaomin Yi is with the School
of Computer Science, Sichuan University, China. \protect 
 E-mail: qiaominy@stu.scu.edu.cn
 
\IEEEcompsocthanksitem Ning Yang is the corresponding author and with the School
of Computer Science, Sichuan University, China. \protect 
 E-mail: yangning@scu.edu.cn

\IEEEcompsocthanksitem Philip S. Yu is with the Department of Computer Science, University of Illinois at Chicago, USA. \protect 
E-mail: psyu@uic.edu
}
\thanks{}}

\markboth{Journal of \LaTeX\ Class Files,~Vol.~14, No.~8, August~2015}%
{Shell \MakeLowercase{\textit{et al.}}: Bare Demo of IEEEtran.cls for Computer Society Journals}

\IEEEcompsoctitleabstractindextext{%
\begin{abstract}

Robust recommendation aims at capturing true preference of users from noisy data, for which there are two lines of methods have been proposed. One is based on noise injection, and the other is to adopt the generative model Variational Auto-encoder (VAE). However, the existing works still face two challenges. First, the noise injection based methods often draw the noise from a fixed noise distribution given in advance, while in real world, the noise distributions of different users and items may differ from each other due to personal behaviors and item usage patterns. Second, the VAE based models are not expressive enough to capture the true preference since VAE often yields an embedding space of a single modal, while in real world, user-item interactions usually exhibit multi-modality on user preference distribution. In this paper, we propose a novel model called Dual Adversarial Variational Embedding (DAVE) for robust recommendation, which can provide personalized noise reduction for different users and items, and capture the multi-modality of the embedding space, by combining the advantages of VAE and adversarial training between the introduced auxiliary discriminators and the variational inference networks. The extensive experiments conducted on real datasets verify the effectiveness of DAVE on robust recommendation. 

\end{abstract}

}

\maketitle

\IEEEdisplaynotcompsoctitleabstractindextext

\IEEEpeerreviewmaketitle

\section{Introduction}
\IEEEPARstart{R}{ecommender} systems have been attracting growing interest of researchers due to their vital role in various online applications, such as e-commerce and social media. In recommender systems, recommendation models are learned from historical interaction (feedback) data which are often seen as noise-free by most of the existing works. In big data era, however, data are usually full of noise. For example, one click of a user might be a random operation which cannot represent the true preference of the user. Noisy data will cause the recommendation models without robustness to weak generalizability and inability to capture true preference from data with even small perturbations \cite{Yuan2019Adversarial}. 

Recently, a few methods have been proposed for robust recommendation, which roughly follow two lines. One line of the existing works improves the robustness of recommendation models by injecting extra noise to training data or model parameters during training process \cite{Yao2016Collaborative,He2018Adversarial,Yuan2019Adversarial,Li2020Adversarial}, while the other line adopts a generative process to learn powerful recommendation models with noise tolerability \cite{Yu2019VAEGAN,liang2018variational,shenbin2020recvae}. However, robust recommendation is still far from being well solved partly due to the following challenges.

\begin{itemize}

\item \textbf{Personalized Noise Reduction} In the line of the existing works that obtain robustness by injecting extra noise to training data or model parameters, the added noise is often drawn from a fixed probability distribution shared by different users \cite{Yao2016Collaborative,He2018Adversarial,Yuan2019Adversarial,Li2020Adversarial}, where the underlying assumption is that data of different users have the same noise level. In real world, however, the noise in data of different users and items has different distribution, due to different behaviors and item usage patterns. Therefore, as part of personalization, robust recommendation is expected to provide personalized noise reduction with adaptability to different noise distributions.

\item \textbf{Multimodal Distribution of Preference} Inspired by the success of Variational Auto-encoder (VAE) in computer vision, one line of the existing works on robust recommendation captures user preference by latent embeddings generated from VAE based models \cite{Li2017Collaborative,liang2018variational,Bai2019Collaborative,shenbin2020recvae}. However, recent studies show that VAE tends to yield a latent space with a single modal that is not expressive enough to capture the true posterior distribution of embeddings \cite{Mescheder2017Adversarial}. In the context of recommender systems, user-item interactions often exhibit multi-modality on user preference distribution, i.e., different users have different preference distributions around different modes. For example, in music recommender systems, users’ preferences to music styles can be separated into multiple clusters each of which can be viewed as a unique distribution, say, a Gaussian distribution with its unique mean (a specific music style) and variance. Such multi-modality means that users’ preferences should be approximated with multiple distributions rather than with only one as the existing works did. Therefore, to improve the robustness of the embedding learning for users and items, we need a model expressive enough to capture the multi-modality of preference.

\end{itemize}

In this paper, to address the above challenges, we propose a novel model called Dual Adversarial Variational Embedding (DAVE) for robust recommendation. The main idea of DAVE is to adaptively capture the different noise distributions of different users or items and the multi-modality of preference with an inferred distribution unique to a user or an item, using variational inference combined with adversarial training. At first, to provide personalized noise reduction for different users and items, DAVE introduces two VAEs to infer a unique latent distribution for each user and item in a dual form, respectively, from which user and item embeddings against noise can be drawn and then fed into a neural collaborative filtering network \cite{He2017Neural} for subsequent preference prediction. Here the advantages are two-fold. The first advantage is that by the power of variational inference, DAVE can adaptively generate a unique embedding distribution for each user and item for their personalized noise reduction, instead of manually setting a fixed noise level. The second advantage is that to enhance the robustness of embeddings, the noise is modeled with the variance of the inferred distributions, which can be viewed as corrupting the latent space by learning the stochastic noise of user-item interactions, unlike the traditional methods where embeddings of users and items are essentially a point estimation. In DAVE, the VAEs are trained jointly with the subsequent neural collaborative filtering network, where the objectives are minimizing the decoding error of VAE and the prediction error of neural collaborative filtering simultaneously. 

To improve the expressiveness of DAVE for capturing the preference multi-modality, inspired by the idea of Adversarial Variational Bayes \cite{Mescheder2017Adversarial}, we further introduce two auxiliary discriminators together with the inference networks of the VAEs to form two Generative Adversarial Networks (GAN), which can regularize the learning of the inference networks of the VAEs for users and items, respectively. Under the framework of GAN, the discriminator, which estimates the probabilities of sampling from the true distribution, and the generator, which captures the underlying data distribution, are jointly trained in an adversarial fashion. In DAVE, the inference network of VAE plays the role of generator. Unlike traditional VAE training where an explicit representation of the posterior distribution is required for the computation of the KL-divergence, the adversarial training between the inference network and the auxiliary discriminator can approximate the KL-divergence regularization for any complex posterior distributions, without the need for explicitly representing the posterior distribution with parametric expression. Such flexibility enables DAVE to capture the multi-modality of user preferences by inferring complex posterior distributions that are multimodal and cannot be explicitly formulated.

The main contributions of this paper are summarized as follows:

\begin{enumerate}[(1)]

\item We propose a novel model called Dual Adversarial Variational Embedding (DAVE) for robust recommendation, which can provide personalized noise reduction for different users and items, and capture the multi-modality of preference. 

\item For the personalized noise reduction, we introduce two VAEs, which are jointly learned with a neural collaborative filtering network, to infer a unique embedding distribution for each user and item, respectively. Due to the variational inference power of VAE, the noise levels of different users or items can be adaptively captured by their own distributions from which robust embeddings can be drawn.

\item To capture the multi-modality of preference, we introduce two auxiliary discriminators for user and item, respectively, to regularize the learning of the inference networks via an adversarial training, which endows DAVE with the flexibility to infer the preference distributions with multi-modality.

\item We conduct extensive experiments on real world datasets and the experimental results verify the effectiveness of the proposed model. 

\end{enumerate}

The rest of this paper is organized as follows. In Section 2, we introduce the preliminaries and formally define the target problem. We present the details of DAVE in Section 3. In Section 4, we empirically evaluate the performance of DAVE over real world datasets, verify the robustness and expressiveness of DAVE, and analyze the influence of hyper-parameters. At last, we briefly review the related works in Section 5 and conclude in Section 6.

\section{Preliminaries and Problem Definition}
\subsection{Notation Definition}
Let $\boldsymbol{U}$ be the set of $N$ users, and $\boldsymbol{V}$ be the set of $M$ items. We define the user-item interaction matrix as $\boldsymbol{R}\in\mathbb{R}^{N\times M}$ based on implicit feedbacks (e.g., clicking, buying, or commenting an item), where $R_{uv}$ = 1 if the interaction between user $u \in \boldsymbol{U}$ and item $v \in \boldsymbol{V}$ is observed, otherwise $R_{uv}$ = 0. We associate each user $u\in\boldsymbol{U}$ with two vectors. One is the user interaction vector $\boldsymbol{u}\in\{0,1\}^M$, which is the transpose of $u$-th row of $\boldsymbol{R}$, and the other is the user embedding $\boldsymbol{x}_{u}\in\mathbb{R}^d$, where $d$ is the dimensionality of user latent representation. Similarly, each item $\mathnormal{v}$$\in$$\boldsymbol{V}$ are also associated with two vectors. One is the item interaction vector $\boldsymbol{v}\in\{0,1\}^N$, which corresponds to the $v$-th column of $\boldsymbol{R}$, and the other is the item embedding $\boldsymbol{y}_v\in\mathbb{R}^d$. The set of items that a user $u$ has interacted with is denoted by $\boldsymbol{V} _{\text{u}}$, i.e., $\boldsymbol{V} _{\text{u}} = \{ v | R_{uv} = 1 \}$. Its complementary set is denoted by $\boldsymbol{\overline{V}}_{\text{u}}$, i.e., $\boldsymbol{\overline{V}}_{\text{u}} = \boldsymbol{V} \setminus \boldsymbol{V}_{\text{u}} = \{ v | R_{uv} = 0 \}$, which is the set of items that $u$ has not interacted with.

\begin{figure*}
	\centering \includegraphics[width=1\textwidth]{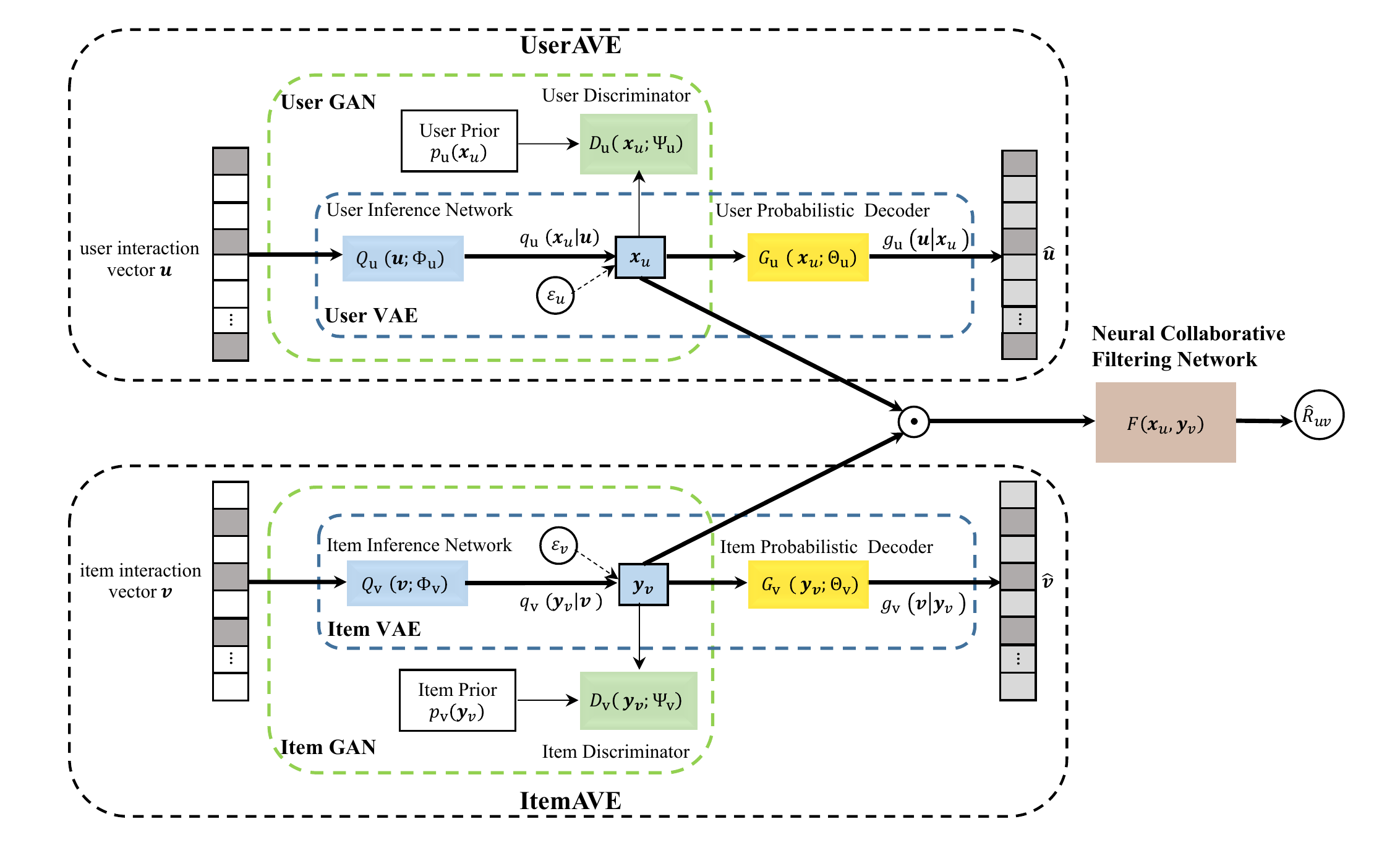}
	\caption{The architecture of DAVE.}
	\label{Fig_DAVE}
\end{figure*}

\subsection{Problem Definition}
Given a user set $\boldsymbol{U}$, an item set $\boldsymbol{V}$, and the observed interaction matrix $\boldsymbol{R}$, our goal is to recommend to a specific user $u \in \boldsymbol{U}$ an item $v \in \boldsymbol{\overline{V}}_{\text{u}}$ with maximal predicted $\widehat{R}_{uv}$. The predicted score $\widehat{R}_{uv}$ is constrained to the range [0, 1], which represents the probability of $u$ will interact with $v$.

\begin{table}[t]
    \centering
    \caption{Notations}
    \begin{tabular}{ll}
    \hline
    Symbol & Description\\
    \hline
    
    $\boldsymbol{u}$ & user interaction vector\\
    $\boldsymbol{v}$ & item interaction vector\\
    $\boldsymbol{x}_u$ & embedding of user $u$\\
    $\boldsymbol{y}_v$ & embedding of item $v$ \\
    $\widehat{\boldsymbol{u}}$ & reconstructed user interaction vector\\
    $\widehat{\boldsymbol{v}}$ & reconstructed item interaction vector\\
    $p_{\text{u}}(\cdot)$ & user prior\\
     $p_{\text{v}}(\cdot)$ & item prior\\
    $q_{\text{u}}(\cdot|\boldsymbol{u})$ & posterior of user embeddings \\
    $q_{\text{v}}(\cdot|\boldsymbol{v})$ & posterior of item embeddings \\
    $g_{\text{u}}(\cdot|\boldsymbol{x}_u)$ & distribution of reconstructed user interaction vectors \\
    $g_{\text{v}}(\cdot|\boldsymbol{y}_v)$ & distribution of reconstructed item interaction vectors \\
    $Q_{\text{u}}(\cdot; \Phi_{\text{u}})$ & user inference network with parameters $\Phi_{\text{u}}$ \\
    $Q_{\text{v}}(\cdot;\Phi_{\text{v}})$ & item inference network with parameters $\Phi_{\text{v}}$ \\
    $D_{\text{u}}(\cdot;\Psi_{\text{u}})$ & user discriminator with parameters $\Psi_{\text{u}}$ \\
    $D_{\text{v}}(\cdot;\Psi_{\text{v}})$ & item discriminator with parameters $\Psi_{\text{v}}$ \\
    $G_{\text{u}}(\cdot; \Theta_{\text{u}})$ & user probabilistic decoder with parameters $\Theta_{\text{u}}$ \\
    $G_{\text{v}}(\cdot;\Theta_{\text{v}})$ & item probabilistic decoder with parameters $\Theta_{\text{v}}$ \\
    $\epsilon_u$ & auxiliary noise for user embedding \\
     $\epsilon_v$ & auxiliary noise for item embedding \\
    \hline
    \end{tabular}
    \label{Tbl_Notation}
\end{table}

\section{The Proposed Method}
In this section, we first present the architecture of the proposed Dural Adversarial Variational Embedding (DAVE) model, and then describe its optimization objective and learning in detail.

\subsection{Architecture of DAVE}
Figure \ref{Fig_DAVE} shows the architecture of DAVE, where the main notations are summarized in Table \ref{Tbl_Notation}. From the Vertical view, we can see that DAVE comprises two dual parts including the User Adversarial Embedding (UserAVE) and the Item Adversarial Embedding (ItemAVE), which respectively take a user interaction vector $\boldsymbol{u}$ and an item interaction vector $\boldsymbol{v}$ as inputs, and generate the user embedding $\boldsymbol{x}_{u}$ and the item embedding $\boldsymbol{y}_v$. Once the user and item embeddings are prepared, DAVE will make the rating prediction $\widehat{R}_{uv}$ by feeding the embeddings into a neural collaborative filtering function $F(\boldsymbol{x}_{u}, \boldsymbol{y}_v)$ which is realized by an MLP network \cite{He2017Neural}. 

In Figure \ref{Fig_DAVE}, the UserAVE consists of three parts, (1) user inference network ($Q_{\text{u}}$), which takes the user interaction vector ($\boldsymbol{u}$) to generate the user embedding ($\boldsymbol{x}_{u}$), (2) user probabilistic decoder ($G_{\text{u}}$), which reconstructs the the user interaction vector ($\widehat{\boldsymbol{u}}$) from the user embedding ($\boldsymbol{x}_{u}$), and (3) the user discriminator ($D_{\text{u}}$) to help the user inference network in (1) to generate more robust user embedding. 

In particular, UserAVE uses a variational inference network $Q_{\text{u}}(\boldsymbol{u}; \Phi_{\text{u}})$ with parameters $\Phi_{\text{u}}$ to infer a unique posterior distribution $q_{\text{u}}(\boldsymbol{x}_{u}|\boldsymbol{u})$ of the latent representation for each user $u$, from which the embedding $\boldsymbol{x}_{u}$ of user $u$ can be drawn out. The inference network $Q_{\text{u}}(\boldsymbol{u}; \Phi_{\text{u}})$ is trained jointly with the probabilistic decoder $G_{\text{u}}(\boldsymbol{x}_{u}; \Theta_{\text{u}})$ with parameters $\Theta_{\text{u}}$, which reconstructs the user interaction vector $\boldsymbol{u}$ as $\widehat{\boldsymbol{u}}$ with the probability $g_{\text{u}}(\boldsymbol{u}|\boldsymbol{x}_{u})$ that minimizes the reconstruction error. Due to the merit of variational inference, UserVAE can capture the different noise distributions of the interaction data for different users by the variance of their unique posterior distribution. Note that the dashed arrow represents sampling auxiliary noise $\epsilon_{u}$ from the standard normal distribution, which will be used for the reparameterization trick \cite{Kingma2014Auto, Rezende2014Stochastic} to optimize the inference network.

As we have mentioned before, due to the nature of VAE, the inferred posterior distributions $q_{\text{u}}(\boldsymbol{x}_{u}|\boldsymbol{u})$ for different users tend to lie around a single mode \cite{Kingma2014Auto}, which might make the embedding spaces of different users indistinguishable from each other and consequently fail to accurately capture the user personalized preference. To improve the expressiveness of the user embeddings, we further introduce an auxiliary discriminator $D_{\text{u}}(\boldsymbol{x}_{u}; \Psi_{\text{u}})$ with parameters $\Psi_{\text{u}}$, which together with the generator $Q_{\text{u}}(\boldsymbol{u}; \Phi_{\text{u}})$ forms a Generative Adversarial Network (GAN). The optimization objective of $D_{\text{u}}(\boldsymbol{x}_{u}; \Psi_{\text{u}})$ is to distinguish the user embeddings drawn from two distributions: the inferred posterior distribution $q_{\text{u}}(\boldsymbol{x}_{u}|\boldsymbol{u})$ unique to each user, and a given prior distribution $p_{\text{u}}(\boldsymbol{x}_{u}) \sim \mathcal{N}(\mathbf{0}, \mathbf{I})$ shared among different users. As we will see in the experiments, the adversarial training of the discriminator $D_{\text{u}}(\boldsymbol{x}_{u}; \Psi_{\text{u}})$ against the generator $Q_{\text{u}}(\boldsymbol{u}; \Phi_{\text{u}})$ can approximate the KL-divergence regularization between any complex posterior distributions and the prior distribution, which offers DAVE the flexibility without the need to make an explicit assumption about the posterior distribution, and enables DAVE to infer complex posterior distributions that are multimodal and cannot be formulated explicitly.

As the dual part, the structure of ItemAVE is similar to that of UserAVE. Particularly, ItemAVE also generates the robust item embeddings $\boldsymbol{y}_v$ via a VAE with item interaction vector $\boldsymbol{v}$ as input, where the variational inference network $Q_{\text{v}}(\boldsymbol{v}; \Phi_{\text{v}})$ with parameters $\Phi_{\text{v}}$, the probabilistic decoder $G_{\text{v}}(\boldsymbol{y}_v; \Theta_V)$ with parameters $\Theta_V$, and the auxiliary noise variable $\epsilon_v$ are the counterparts of $Q_{\text{u}}(\boldsymbol{u}; \Phi_{\text{u}})$, $G_{\text{u}}(\boldsymbol{x}_{u}; \Theta_{\text{u}})$, and $\epsilon_{u}$ in UserAVE. Again due to the variational inference power, the item embeddings will also benefit from the item posterior distribution $q_{\text{v}}(\boldsymbol{y}_v | \boldsymbol{v})$ unique to each item $v$, which makes them adaptable to the different noise distributions in interaction data of different items. Similar to UserAVE, ItemAVE introduces an auxiliary discriminator $D_{\text{v}}(\boldsymbol{y}_v; \Psi_{\text{v}})$ with parameters $\Psi_{\text{v}}$, of which the role is also to distinguish the item embeddings drawn from each inferred unique posterior distributions $q_{\text{v}}(\boldsymbol{y}_v | \boldsymbol{v})$ from those drawn from a given prior distribution $p_{\text{v}}(\boldsymbol{y}_v)  \sim \mathcal{N}(\mathbf{0}, \mathbf{I})$. Similar to UserAVE, the multi-modality of the inferred posterior embedding distributions of different items can also be captured due to the flexibility offered by the adversarial training of the $D_{\text{v}}(\boldsymbol{y}_v; \Psi_{\text{v}})$ and $Q_{\text{v}}(\boldsymbol{v}; \Phi_{\text{v}})$.

\subsection{Objective Function}
As UserAVE, ItemAVE, and the neural collaborative filtering network $F(\boldsymbol{x}_{u}, \boldsymbol{y}_v)$ will be jointly trained in an end-to-end fashion, the overall objective function of DAVE is defined as
\begin{equation}
\begin{split}
\mathcal{L} = \mathcal{L}_{\text{u}}+\mathcal{L}_{\text{v}}+\mathcal{L}_{\text{f}}, 
\end{split}
\end{equation}
where $\mathcal{L}_{\text{u}}$, $\mathcal{L}_{\text{v}}$, and $\mathcal{L}_{\text{f}}$ are the objective functions of UserAVE, ItemAVE, and $F(\boldsymbol{x}_{u}, \boldsymbol{y}_v)$, respectively, which will be detailed in the following subsections.

\subsubsection{Objective Function of UserAVE}

As we have mentioned early, UserAVE consists of a VAE and an auxiliary GAN, and therefore its optimization objective is 
\begin{equation}
\mathcal{L}_{\text{u}} = \mathcal{L}_{\text{u}}^{\text{VAE}} + \mathcal{L}_{\text{u}}^{\text{D}},
\end{equation}
where $\mathcal{L}_{\text{u}}^{\text{VAE}}$ and $\mathcal{L}_{\text{u}}^{\text{D}}$ are the objective functions of VAE and the auxiliary GAN in UserAVE, respectively. 

Traditional objective function of VAE usually regularizes the variational inference network $Q_{\text{u}}$ using KL-divergence, which cannot serve our purpose to learn multimodal embedding space as it will cause the inferred posterior distributions $q_{\text{u}}(\boldsymbol{x}_{u}|\boldsymbol{u})$ indistinguishable for different $\boldsymbol{u}$. Our overall idea to overcome this issue is to instead regularize $Q_{\text{u}}$ via an adversarial training between it and the auxiliary discriminator $D_{\text{u}}$. Due to such adversarial regularization, the inferred posterior distributions $q_{\text{u}}(\boldsymbol{x}_{u}|\boldsymbol{u})$ of different users can stay away from each other, which benefits the learning of multimodal embedding space.

As the observed data are just the user interaction vectors $\boldsymbol{u}$, we will learn the parameters of VAE, $\Phi_{\text{u}}$ and $\Theta_{\text{u}}$, by maximizing the log-likelihood $\log p(\boldsymbol{u})$, where $p(\boldsymbol{u})$ is the distribution of user interaction data. By applying Variational Bayes and Jensen's inequality \cite{Kingma2014Auto}, for a specific user $u$ we have
\begin{equation}
\begin{split}
\log p(\boldsymbol{u})\geq\mathbb E_{\boldsymbol{x} _{u}\sim q_{\text{u}}(\boldsymbol{x}_{u} | \boldsymbol{u})} & [\log g_{\text{u}}(\boldsymbol{u} | \boldsymbol{x}_{u})] \\ 
& - \text{KL}(q_{\text{u}}(\boldsymbol{x} _{u} | \boldsymbol{u}) \parallel p_{\text{u}}(\boldsymbol{x_{\text{u}}})),  
\end{split}
\end{equation}
where the right side is the evidence lower bound (ELBO) \cite{Blei2017Variational} also known as variational lower bound, and $p_{\text{u}}(\boldsymbol{x_{\text{u}}})$ is the Gaussian prior of $\boldsymbol{x}_{u}$. In the ELBO, the variational distribution $q_{\text{u}}(\boldsymbol{x} _{u} | \boldsymbol{u})$ is also a Gaussian distribution with mean $\boldsymbol{\mu}_{u}$ and variance $\boldsymbol{\sigma}_{\text{u}}^2$, which are outputs of the inference network $q_{\text{u}}(\boldsymbol{u};\Phi_{\text{u}})$, and $g_{\text{u}}(\boldsymbol{u} | \boldsymbol{x}_{u})$ is the reconstruction probability dependent on the decoder $G_{\text{u}}(\Theta_{\text{u}})$. In variational inference, maximizing the log-likelihood is reduced to the maximization of ELBO over user set $\boldsymbol{U}$, i.e., 

\begin{equation}
\begin{split}
\mathcal{L}_{\text{u}}^{\text{VAE}}( \Theta_{\text{u}}, \Phi_{\text{u}})=\mathbb{E}_{\boldsymbol{u} \sim p(\boldsymbol{u})}&\Big[\mathbb E_{\boldsymbol{x}_{u}\sim q_{\text{u}}(\boldsymbol{x}_{u}\mid\boldsymbol{u})}[\log g_{\text{u}}(\boldsymbol{u} \mid \boldsymbol{x} _{u})]\\&- \text{KL}(q_{\text{u}}(\boldsymbol{x} _{u} \mid \boldsymbol{u}) \parallel p_{\text{u}}(\boldsymbol{x}_{u})) \Big].
\end{split}
\label{L_U_VAE}
\end{equation}
Note that the reconstruction probability $g_{\text{u}}(\boldsymbol{u} \mid \boldsymbol{x} _{u})$ implicitly represents the negative reconstruction error \cite{Kingma2014Auto}, and therefore, maximizing its expectation is equivalent to minimizing the expected reconstruction error. 

As the KL-divergence is
\begin{equation}
\mathbb{E}_{\boldsymbol{x}_{u}\sim q_{\text{u}}(\boldsymbol{x}_{u}\mid\boldsymbol{u})} [\log q_{\text{u}}(\boldsymbol{x} _{u} | \boldsymbol{u}) - \log p_{\text{u}}(\boldsymbol{x}_{u})],
\label{Eq_KL}
\end{equation}
we can rewrite the objective function (\ref{L_U_VAE}) as
\begin{equation}
\begin{split}
\mathcal{L}_{\text{u}}^{\text{VAE}}( \Theta_{\text{u}}, \Phi_{\text{u}})=\mathbb{E}_{\boldsymbol{u} \sim p(\boldsymbol{u})} &\mathbb E_{\boldsymbol{x}_{u}\sim q_{\text{u}}(\boldsymbol{x}_{u}\mid\boldsymbol{u})} \big[\log g_{\text{u}}(\boldsymbol{u} \mid \boldsymbol{x} _{u}) \\& + \log p_{\text{u}}(\boldsymbol{x}_{u}) - \log q_{\text{u}}(\boldsymbol{x} _{u} | \boldsymbol{u}) \big].
\end{split}
\label{L_U_VAE_2}
\end{equation}

When the variational posterior distribution $q_{\text{u}}(\boldsymbol{x} _{u} | \boldsymbol{u})$ has explicit representation that is tractable like Gaussian distribution, we can directly compute the objective function for maximization. However, recent studies \cite{Ilya2017Wasserstein,Mescheder2017Adversarial,Husz2017Variational} show that if the VAE is optimized only according to Equation (\ref{L_U_VAE_2}), the KL-divergence will make the variational posterior distributions $q_{\text{u}}(\boldsymbol{x} _{u} | \boldsymbol{u})$ of different users all close to the same prior $p_{\text{u}}(\boldsymbol{x}_{u})$, which results in inferior user embeddings that are not expressive enough to capture the multi-modality of the preference distributions of users. Therefore, to improve the expressiveness of the variational inference network of UserAVE, we assume $q_{\text{u}}(\boldsymbol{x} _{u} | \boldsymbol{u})$ is implicit and introduce an auxiliary discriminator $D_{\text{u}}(\boldsymbol{x}_{u}; \Psi_{\text{u}})$ together with the inference network $Q_{\text{u}}(\boldsymbol{u}; \Phi_{\text{u}})$ as generator to form a GAN to help the inference of the implicit posterior distribution $q_{\text{u}}(\boldsymbol{x} _{u} | \boldsymbol{u})$. It is easy to define the objective function of the discriminator as
\begin{equation}
\begin{split}
\mathcal{L}_{\text{u}}^{\text{D}}(\Psi_{\text{u}}) &=\mathbb{E}_{\boldsymbol{u} \sim p_(\boldsymbol{u})} \mathbb{E}_{\boldsymbol{x} _{u}\sim p_{\text{u}}(\boldsymbol{x} _{u})}\log\sigma(D_{\text{u}}(\boldsymbol{x} _{u}; \Psi_{\text{u}}))\\&+\mathbb{E}_{\boldsymbol{u}\sim p(\boldsymbol{u})}\mathbb{E}_{\boldsymbol{x} _{u}\sim q_{\text{u}}(\boldsymbol{x} _{u} \mid\boldsymbol{u})} \log(1-\sigma(D_{\text{u}}(\boldsymbol{x} _{u}; \Psi_{\text{u}})),
\end{split}
\label{L_U_GAN}
\end{equation}
where $\sigma(\cdot)$ is the sigmoid function. As we treat the samples from the prior as true while ones from the variational posterior as fake, it is easy to show that when the generator $Q_{\text{u}}(\boldsymbol{u}; \Phi_{\text{u}})$ is fixed (and equivalently, $q_{\text{u}}(\boldsymbol{x} _{u} \mid\boldsymbol{u})$ is fixed), $\mathcal{L}_{\text{u}}^{\text{D}}$ achieves its maximum at $\Psi^*_{\text{u}}$ such that 
\begin{equation}
D_{\text{u}}(\boldsymbol{x}_{u}; \Psi^*_{\text{u}})=\log p_{\text{u}}(\boldsymbol{x}_{u}) - \log q_{\text{u}}(\boldsymbol{x}_{u}\mid\boldsymbol{u}), 
\end{equation}
where $\Psi^*_{\text{u}}$ represents the optimal parameters of the discriminator \cite{goodfellow2014generative}. Therefore, Equation (\ref{L_U_VAE_2}) can be further rewritten as
\begin{equation}
\begin{split}
\mathcal{L}_{\text{u}}^{\text{VAE}}( \Theta_{\text{u}}, \Phi_{\text{u}})=\mathbb{E}_{\boldsymbol{u} \sim p(\boldsymbol{u})} &\mathbb E_{\boldsymbol{x}_{u}\sim q_{\text{u}}(\boldsymbol{x}_{u}\mid\boldsymbol{u})} \big[\log g_{\text{u}}(\boldsymbol{u} \mid \boldsymbol{x} _{u}) \\& +  D_{\text{u}}(\boldsymbol{x}_{u}; \Psi^*_{\text{u}}) \big].
\end{split}
\label{L_U_VAE_3}
\end{equation}

As we have mentioned before, the insight here is that although the adversarial training between $D_{\text{u}}$ and $Q_{\text{u}}$ will also cause the posterior distributions inferred by $Q_{\text{u}}$ to be close to the prior, which exactly approximates the regularization that minimizes the KL-divergence of the posterior distribution to the prior, it obviates the requirement to explicitly represent the posterior distribution. This is in contrast with the existing VAE based models. Such flexibility enables DAVE to infer any complex posterior distributions that are multimodal and cannot be formulated with a simple parametric expression, with the result that the drawn user embeddings have chance to stay far away from each other, which benefits the capturing of diverse user preference distributions.

The user embeddings $\boldsymbol{x}_u$ are generated by sampling from $q_{\text{u}}(\boldsymbol{x} _{u} | \boldsymbol{u})$, which, however, makes the objective functions not differentiable. Using the reparameterization trick \cite{Kingma2014Auto}, we draw an auxiliary noise $\boldsymbol{\epsilon}_{u} \sim \mathcal{N}(\mathbf{0}, \mathbf{I})$ and instead generate the user embeddings by a composite function $\phi_{\text{u}}(\boldsymbol{u}, \boldsymbol{\epsilon}_{u}) = \phi_{\text{u}}(Q_{\text{u}}(\boldsymbol{u}; \Phi_{\text{u}}), \boldsymbol{\epsilon}_{u})$. Then Equations (\ref{L_U_GAN}) and (\ref{L_U_VAE_3}) can be finally rewritten as follows:
\begin{equation}
\begin{split}
\mathcal{L}_{\text{u}}^{\text{D}}&(\Psi_{\text{u}}) =\mathbb{E}_{\boldsymbol{u} \sim p_(\boldsymbol{u})} \mathbb{E}_{\boldsymbol{x} _{u}\sim p_{\text{u}}(\boldsymbol{x} _{u})}\log\sigma(D_{\text{u}}(\boldsymbol{x} _{u}; \Psi_{\text{u}}))\\& + \mathbb{E}_{\boldsymbol{u}\sim p(\boldsymbol{u})}\mathbb{E}_{\boldsymbol{\epsilon}_{u} \sim \mathcal{N}(\mathbf{0}, \mathbf{I})} \log(1-\sigma(D_{\text{u}}(\phi_{\text{u}}(\boldsymbol{u}, \boldsymbol{\epsilon}_{u}); \Psi_{\text{u}})),
\end{split}
\label{L_U_GAN_2}
\end{equation}
and
\begin{equation}
\begin{split}
\mathcal{L}_{\text{u}}^{\text{VAE}}( \Theta_{\text{u}}, \Phi_{\text{u}})=\mathbb{E}_{\boldsymbol{u} \sim p(\boldsymbol{u})} &\mathbb E_{\boldsymbol{\epsilon}_{u} \sim \mathcal{N}(\mathbf{0}, \mathbf{I})} \big[\log g_{\text{u}}(\boldsymbol{u} \mid \phi_{\text{u}}(\boldsymbol{u}, \boldsymbol{\epsilon}_{u}) ) \\& + D_{\text{u}}(\phi_{\text{u}}(\boldsymbol{u}, \boldsymbol{\epsilon}_{u}) ; \Psi^*_{\text{u}}) \big].
\end{split}
\label{L_U_VAE_4}
\end{equation}
In the experiments of this paper, we define the reparameterization function as $\phi_{\text{u}}(\boldsymbol{u}, \boldsymbol{\epsilon}_{u}) = \boldsymbol{\mu}_{u}+\boldsymbol{\sigma}_{u}\odot\boldsymbol{\epsilon}_{u}$, where $\odot$ is element-wise product, $\boldsymbol{\mu}_{u}$ and $\boldsymbol{\sigma}^2_{u}$ are the mean and variance of $q_{\text{u}}(\boldsymbol{x} _{u} | \boldsymbol{u})$, respectively. Note that $\boldsymbol{\mu}_{u}$ and $\boldsymbol{\sigma}^2_{u}$ are outputs of the inference network $Q_{\text{u}}(\boldsymbol{u}; \Phi_{\text{u}})$.

\subsubsection{Objective Function of ItemAVE}
As ItemAVE is the dual part of UserAVE, its objective function can be similarly defined as
\begin{equation}
\mathcal{L}_{\text{v}} = \mathcal{L}_{\text{v}}^{\text{VAE}} + \mathcal{L}_{\text{v}}^{\text{D}},
\end{equation}
where $\mathcal{L}_{\text{v}}^{\text{VAE}}$ and $\mathcal{L}_{\text{v}}^{\text{D}}$ are the objective functions of VAE and the auxiliary GAN in ItemAVE, respectively. Through a similar derivation, $\mathcal{L}_{\text{v}}^{\text{VAE}}$ and $\mathcal{L}_{\text{v}}^{\text{D}}$ can be respectively defined as follows:
\begin{equation}
\begin{split}
\mathcal{L}_{\text{v}}^{\text{D}}&(\Psi_{\text{v}}) = \mathbb{E}_{\boldsymbol{v} \sim p_(\boldsymbol{v})} \mathbb{E}_{\boldsymbol{y} _{v}\sim p_{\text{v}}(\boldsymbol{y} _{v})}\log\sigma(D_{\text{v}}(\boldsymbol{y} _{v}; \Psi_{\text{v}}))\\&+\mathbb{E}_{\boldsymbol{v}\sim p(\boldsymbol{v})}\mathbb{E}_{\boldsymbol{y} _{v}\sim q_{\text{v}}(\boldsymbol{y} _{v} \mid\boldsymbol{v})} \log(1-\sigma(D_{\text{v}}(\phi_{\text{v}}(\boldsymbol{v}, \boldsymbol{\epsilon}_{v}); \Psi_{\text{v}}))
\end{split}
\label{L_V_GAN}
\end{equation}
and
\begin{equation}
\begin{split}
\mathcal{L}_{\text{v}}^{\text{VAE}}( \Theta_{\text{v}}, \Phi_{\text{v}})=\mathbb{E}_{\boldsymbol{v} \sim p(\boldsymbol{v})} &\mathbb E_{\boldsymbol{\epsilon}_{v} \sim \mathcal{N}(\mathbf{0}, \mathbf{I})} \big[\log g_{\text{v}}(\boldsymbol{v} \mid \phi_{\text{v}}(\boldsymbol{v}, \boldsymbol{\epsilon}_{v}) ) \\& + D_{\text{v}}(\phi_{\text{v}}(\boldsymbol{v}, \boldsymbol{\epsilon}_{v}) ; \Psi^*_{\text{v}}) \big],
\end{split}
\label{L_V_VAE_4}
\end{equation}
where auxiliary noise $\boldsymbol{\epsilon}_{v} \sim \mathcal{N}(\mathbf{0}, \mathbf{I})$ and $\Psi^*_{\text{v}}$ is the optimal parameters of the discriminator $D_{\text{v}}(\boldsymbol{y} _{v}; \Psi_{\text{v}})$. The reparameterization function is defined as $\phi_{\text{v}}(\boldsymbol{v}, \boldsymbol{\epsilon}_{v}) = \boldsymbol{\mu}_{v}+\boldsymbol{\sigma}_{v}\odot\boldsymbol{\epsilon}_{v}$, where $\boldsymbol{\mu}_{v}$ and $\boldsymbol{\sigma}^2_{v}$ are the mean and variance of $q_{\text{v}}(\boldsymbol{y} _{v} | \boldsymbol{v})$, respectively, which are outputs of the inference network $Q_{\text{v}}(\boldsymbol{v}; \Phi_{\text{v}})$ of ItemAVE.

\subsubsection {Objective Function of Prediction}
For a pair of user $u$ and $v$, once their embeddings $\boldsymbol{x}_{u}$ and $\boldsymbol{y}_{v}$ are generated, DAVE will predict a score $\widehat{R}_{uv}$ of $v$ given by $u$, by feeding the embeddings into the neural collaborative filtering function $F(\boldsymbol{x}_{u}, \boldsymbol{y}_{v})$ which is implemented as the following MLP network \cite{He2017Neural}:
\begin{equation}
\begin{split}
&\boldsymbol{h}_{uv}= a_{L-1}(\boldsymbol{W}_{L-1}(...a_1(\boldsymbol{W}_1(\boldsymbol{x}_{u}\odot \boldsymbol{y}_v)+\boldsymbol{b}_1)...)+\boldsymbol{b}_{L-1})\\
&\widehat{R}_{uv}= a_{L}(\boldsymbol{W}_{L}\boldsymbol{h}_{uv}+\boldsymbol{b}_{L}),
\end{split}
\end{equation}
where $L$ is the number of layers, $\boldsymbol{W}_{i}$, $\boldsymbol{b}_{i}$, $a_{i}$ ($1 \le i \le L$) denote the weight matrix, bias vector, and activation function of the $i$-th layer, respectively. In this paper, we choose ReLU for the activation functions $a_{i}$ ($1 \le i \le L-1$) of the hidden layers, and Sigmoid function for the activation function $a_L$ of the output layer.

We reduce the score prediction to a binary classification over implicit feedback matrix $\boldsymbol{R}$, of which the optimization objective can be defined as to maximize the following log-likelihood function:
\begin{equation}
\begin{split}
\mathcal{L}_{\text{f}}(\Omega) = \sum_{u \in \boldsymbol{U}, v \in \boldsymbol{V}}R_{uv}\log\widehat{R}_{uv}+(1-R_{uv})\log(1-\widehat{R}_{uv}),
\end{split}
\label{L_F}
\end{equation}
where $\Omega = \{  \boldsymbol{W}_{i}, \boldsymbol{b}_{i}, 1 \le i \le L \}$ is the parameters that need to be learned. It is easy to show that maximizing the likelihood $\mathcal{L}_{\text{f}}(\Omega)$ equivalently minimizes the classification error.

\renewcommand{\algorithmicrequire}{\textbf{Input:}}
\renewcommand{\algorithmicensure}{\textbf{Output:}}
\begin{algorithm}[t]
	\caption{Learning DAVE.}
	\label{alg:A}
	\begin{algorithmic}[1]
		\REQUIRE ~~ \\
		User-item interaction matrix $\boldsymbol{R}$, batchsize $B$, dimensionality of embedding $d$.
		\ENSURE ~~ \\
		DAVE parameters $\Phi_{\text{u}},\Phi_{\text{v}},\Theta_{\text{u}},\Theta_{\text{v}},\Psi_{\text{u}},\Psi_{\text{v}}, \Omega$.
		
		\STATE Initialize the parameters.
		\REPEAT
		\STATE Sample a mini-batch $\{(\boldsymbol{u}, \boldsymbol{v})\}$ of size $B$.
		\STATE Fixing $Q_{\text{u}}$, $G_{\text{u}}$,  $Q_{\text{v}}$, $G_{\text{v}}$, and $F$, generate the uer embedding $\boldsymbol{x}_{u}$ and the item embedding $\boldsymbol{y}_v$ for each pair $(\boldsymbol{u}, \boldsymbol{v})$ in the mini-batch by inference networks $Q_{\text{u}}$ and $Q_{\text{v}}$, respectively.
		\STATE For each $\boldsymbol{u}$, sample a $d$-dimensional vectors $\boldsymbol{x}'_{u}$ from user prior $p_{\text{u}}(\boldsymbol{x}_{u})$.
		\STATE For each $\boldsymbol{v}$, sample a $d$-dimensional vectors $\boldsymbol{y}'_{v}$ from item prior $p_{\text{v}}(\boldsymbol{y}_v)$.
		\STATE Update the parameters $\Psi_{\text{u}}$ of $D_{\text{u}}$ with the gradient of $\mathcal{L}_{\text{u}}^{\text{D}}(\Psi_{\text{u}})$ (Equation (\ref{L_U_GAN_2})), using $\{\boldsymbol{x}_{u}\}$ as fake examples and $\{\boldsymbol{x}'_{u}\}$ as real examples.
		\STATE Update the parameters $\Psi_{\text{v}}$ of $D_{\text{v}}$, using the gradient of $\mathcal{L}_{\text{v}}^{\text{D}}(\Psi_{\text{u}})$ (Equation (\ref{L_V_GAN})), using $\{\boldsymbol{y}_{v}\}$ as fake examples and $\{\boldsymbol{y}'_{v}\}$ as real examples.
		\STATE Fixing $D_{\text{u}}$ and $D_{\text{v}}$, jointly update $\Phi_{\text{u}},\Phi_{\text{v}},\Theta_{\text{u}},\Theta_{\text{v}}$, and $\Omega$, with the gradient of the sum of $\mathcal{L}_{\text{u}}^{\text{VAE}}$ (Equation (\ref{L_U_VAE_4})), $\mathcal{L}_{\text{v}}^{\text{VAE}}$ (Equation (\ref{L_V_VAE_4})), and $\mathcal{L}_{\text{f}}$ (Equation (\ref{L_F}).
		\UNTIL convergence.
	\end{algorithmic}
\end{algorithm}

\subsection{Model Learning}
DAVE will be trained with the following objective
\begin{equation}
\max_{\Phi_{\text{u}},\Phi_{\text{v}},\Theta_{\text{u}},\Theta_{\text{v}},\Psi_{\text{u}},\Psi_{\text{v}}, \Omega} \mathcal{L},
\end{equation}
where $\mathcal{L} = \mathcal{L}_{\text{u}}^{\text{VAE}} + \mathcal{L}_{\text{u}}^{\text{D}} + \mathcal{L}_{\text{v}}^{\text{VAE}} + \mathcal{L}_{\text{v}}^{\text{D}} + \mathcal{L}_{\text{f}}$.

To fulfill the adversarial training, the overall training process consists of the following two alternate steps:
\begin{itemize}
	\item Step 1: Fixing $Q_{\text{u}}$, $G_{\text{u}}$, $Q_{\text{v}}$, $G_{\text{v}}$, and the neural collaborative filtering network $F$, optimizing $D_{\text{u}}$ and $D_{\text{v}}$ with respect to $\mathcal{L}^{\text{D}}_{\text{u}}$ (Equation (\ref{L_U_GAN_2})) and $\mathcal{L}^{\text{D}}_{\text{v}}$ (Equation (\ref{L_V_GAN})), respectively;
	
	\item Step 2: Fixing $D_{\text{u}}$ and $D_{\text{v}}$, jointly training $(Q_{\text{u}},G_{\text{u}})$, $(Q_{\text{v}}, G_{\text{v}})$, and $F$, with respect to $\mathcal{L}_{\text{u}}^{\text{VAE}}$ (Equation (\ref{L_U_VAE_4})), $\mathcal{L}_{\text{v}}^{\text{VAE}}$ (Equation (\ref{L_V_VAE_4})), and $\mathcal{L}_{\text{f}}$ (Equation (\ref{L_F})), respectively.
\end{itemize}
Note that at each iteration, the discriminators $D_{\text{u}}$ and $D_{\text{v}}$ should be updated before the variational inference networks $Q_{\text{u}}$, $Q_{\text{v}}$, because $\mathcal{L}_{\text{u}}^{\text{VAE}}$ and $\mathcal{L}_{\text{v}}^{\text{VAE}}$ depend on the optimal $D_{\text{u}}$ and $D_{\text{v}}$ so far. It is also worth noting that in Step 2, the two VAEs,  $(Q_{\text{u}},G_{\text{u}})$ and $(Q_{\text{v}}, G_{\text{v}})$, are trained jointly with the neural collaborative filtering network, by which the reconstruction error and prediction error can be minimized simultaneously. The joint training can take advantage of multi-task learning which makes the supervision signal of $R_{uv}$ able to be propagated back to the inference networks $Q_{\text{u}}$ and $Q_{\text{v}}$. The training procedure is summarized in Algorithm 1 which iteratively updates the parameters of DAVE using mini-batch stochastic gradient ascent.

\section{Experiments}
The experiments mainly aim to answer the following research questions:
\begin{enumerate}[\textbf{RQ}1]
\item How does DAVE perform as compared with state-of-the-art recommendation methods?
\item How is the robustness of DAVE?
\item How is the expressiveness of DAVE?
\item How do the hyper-parameters, embedding dimensionality and negative sampling ratio, affect the performance of DAVE?
\end{enumerate}
Since we use implicit feedback data, as most of the existing work did \cite{He2017Neural, Yao2016Collaborative, Yuan2019Adversarial, He2018Adversarial}, we will evaluate DAVE over top-$k$ recommendation task.

\subsection{Experimental Setting}
\subsubsection{Datasets}
We conduct experiments on five publicly available datasets: Yelp $\footnote{https://github.com/hexiangnan/sigir16-eals}$, Digital Music $\footnote{https://nijianmo.github.io/amazon/index.html}$, MovieLens 1M $\footnote{https://github.com/hexiangnan/neural\_collaborative\_filtering}$ (ML-1M), MovieLens 100K $\footnote{https://grouplens.org/datasets/MovieLens/100k/}$ (ML-100k) and Pinterest $\footnote{https://github.com/hexiangnan/neural\_collaborative\_filtering}$, which are summarized in Table \ref{Tbl_Datasets}. The first four datasets provide users’ explicit ratings on items, so we transform them into implicit data, where each entry is marked as 1 if the rating is observed, otherwise 0. Specially, there are at least 20 ratings for each user in the two MovieLens datasets, while in Yelp, we only retain the users who have at least 10 interactions, due to the higher sparsity of Yelp. In Yelp, a user may rate an item many times. These repetitive ratings count only once in the building of the interaction matrix, which can prevent an interaction from appearing in both the training set and the testing set. Digital Music is a public dataset collected from Amazon. Since it is highly sparse, we only retain the users and items with at lest 5 ratings, which results in a subset that contains 9,906 users and 12,381 items. Pinterest is a dataset consisting of implicit feedbacks, which has been used to evaluate collaborative recommendations on images \cite{He2017Neural, He2018Adversarial}. In Pinterest, an interaction represents a user has pinned an image to his/her board.

\subsubsection{Evaluation Protocol}

To evaluate the performance of DAVE, we adopt the leave-one-out method, which is widely used in top-$k$ recommendation evaluation \cite{Yuan2019Adversarial, He2017Neural, He2018Adversarial}. Specifically, in Yelp, MovieLens 1M and MovieLens 100K, for each user in a dataset, we leave out the latest user-item interaction to form the testing set and use the remaining interactions to form the training set. In Digital Music and Pinterest, since each rating or pin has no timestamp, we randomly leave out one user-item interaction for each user to form the testing set. Note that we also randomly set aside one interaction for each user to form the validation set for the tuning of hyper-parameters. Since it is too time-consuming to rank all items for every user during testing, we follow the common strategy \cite{Yuan2019Adversarial, He2017Neural} that we will check whether the testing item, which the user rated, is ranked ahead of 99 unrated items which are randomly selected from the datasets in advance. The performance of the ranked list is judged by Hit Ratio (HR) and Normalized Discounted Cumulative Gain (NDCG). The HR@$k$ is the ratio of the ranking list that the testing item is ranked in the first $k$ positions, while the NDCG accounts for the position of the hit, which assigns higher weight to hits at higher positions. For both metrics, larger values indicate better performance. We will report the results of $k = 5, 10, 20$ on five datasets.

\begin{table}
	\centering
	\caption{Statistics of Datasets}
	\begin{tabular}{lcccc}
		\toprule
		Dataset&\#Interactions&\#Items&\#Users&Sparsity\\
		\midrule
		Yelp & 730,790 & 25,815 & 25,677 & 99.89\%\\
		Digital Music & 123,518 & 12,381 & 9,906 & 99.90\%\\
		MovieLens 1M& 1,000,209 & 3,706 & 6,040 & 95.53\%\\
		MovieLens 100K & 100,000& 1,682 & 943 & 93.69\%\\
		Pinterest & 1,500,809& 9,916 & 55,187 & 99.73\%\\
		\bottomrule
	\end{tabular}
\label{Tbl_Datasets}
\end{table}

\begin{table*}
\label{Tbl_BS}
	\centering
	\setlength{\tabcolsep}{0.7pt}{
		\caption{Comparison of Baselines}
		\begin{tabular}{p{3.9cm}|p{1.8cm}<{\centering}p{1.5cm}<{\centering}p{1.5cm}<{\centering}p{2.5cm}<{\centering}p{2.0cm}<{\centering}p{1.5cm}<{\centering}p{1.5cm}<{\centering}}
			\hline
			\diagbox{\textbf{Baselines}}{\textbf{Characteristics}}&{User Embedding}&Item Embedding&Noise-tolerant&Personalized noise reduction&Embedding Distribution&Multi-Modality\\
			\hline
			\rule{0pt}{11pt}CDAE& \checkmark &   & \checkmark &  &  &  \\
			\rule{0pt}{11pt}APR& \checkmark & \checkmark & \checkmark &   &  &  \\
			\rule{0pt}{11pt}ACAE& \checkmark &   & \checkmark &   &  &  \\
			\rule{0pt}{11pt}NeuMF & \checkmark & \checkmark &   &  &  &   \\
			\rule{0pt}{11pt}CFGAN& \checkmark &   &   &    &   &  \\
			\rule{0pt}{11pt}AVB& \checkmark&   &   &  & &\checkmark\\
			\rule{0pt}{11pt}VAEGAN& \checkmark&   &   &  & &\checkmark\\
			\rule{0pt}{11pt}CVAE-GAN& \checkmark&   & \checkmark & \checkmark & \checkmark &   \\
			\rule{0pt}{11pt}RecVAE& \checkmark&   & \checkmark & \checkmark & \checkmark &   \\
			\rule{0pt}{11pt}DAVE-adv& \checkmark& \checkmark & \checkmark & \checkmark&\checkmark& \\
			\rule{0pt}{11pt}DAVE+aae& \checkmark& \checkmark & \checkmark &  & & \\
			\rule{0pt}{11pt}DAVE& \checkmark& \checkmark & \checkmark & \checkmark&\checkmark&\checkmark\\
			\bottomrule
	\end{tabular}}
	
\end{table*}

\subsubsection{Baselines}
We will compare DAVE with the following advanced methods, whose characteristics are shown in Table 3.
\begin{itemize}
	\item NeuMF \cite{He2017Neural}: NeuMF is a general framework NCF for collaborative filtering based on neural networks. It employs a Multi-Layer Perceptron (MLP) to model non-linear user-item interactions between latent features of users and items. 
	
	\item CDAE \cite{Yao2016Collaborative}: CADE is a Denoising Auto-encoder based collaborative filtering framework for top-$k$ recommendation. By utilizing denoising technique, CDAE can learn robust latent representations of corrupted user-item interactions for recommendation. 
	
	\item CFGAN \cite{Chae2018CFGAN}: CFGAN is a GAN-based collaborative filtering framework, where a real value vector-wise adversarial training is introduced to improve the representation learning of the users or items.  
	
	\item APR \cite{He2018Adversarial}: APR is an Adversarial Personalized Ranking framework, which enhances the pairwise ranking method BPR \cite{Rendle2012BPR}  by adversarial training. Particularly, APR offers the robustness at the level of model parameters rather than model input, by injecting adversarial noise to parameters of BPR model \cite{Rendle2012BPR} during the adversarial training. 
	
	\item ACAE \cite{Yuan2019Adversarial}: ACAE is a general adversarial training framework for neural network-based recommendation models, which also applies adversarial training for improving the robustness of recommendations. 
	
	\item AVB \cite{Mescheder2017Adversarial}: AVB is a technique for training variational auto-encoders with arbitrarily expressive inference models based on adversarial training, which introduces an auxiliary discriminative network that allows to rephrase the maximum-likelihood problem as a two-player game.

	\item VAEGAN \cite{Yu2019VAEGAN}: VAEGAN is a Collaborative Filtering Framework based on Adversarial Variational auto-encoders, which utilizes a flexible black-box inference model as well as adversarial training to train VAEs for implicit variational inference.
	
	\item CVAE-GAN  \cite{Bao2017CVAE-GAN}: CVAE-GAN is variational generative adversarial networks, which is a general learning framework that combines a variational auto-encoder with a generative adversarial network.
	
	\item RecVAE \cite{shenbin2020recvae}: RecVAE is a Recommender VAE model with a new architecture for the encoder network that can be trained with corrupted implicit user-item interaction vectors.

\end{itemize}

\subsubsection{Parameter Setting}

The hyper-parameters are tuned on validation sets. We set the batch size to 256 for MovieLens 100K, MovieLens 1M and Pinterest, and 128 for Yelp and Digital Music. The negative sampling ratio is set to 4 for Yelp and MovieLens 100K, 3 for MovieLens 1M, and 2 for Digital Music and Pinterest. The embedding dimensionality is set to 32 for Yelp, and 64 for MovieLens 1M, MovieLens 100K, Digsital Music and Pinterest. We use Adam to learn the VAEs $(Q_{\text{u}},G_{\text{u}})$ and $(Q_{\text{v}}, G_{\text{v}})$ and the neural collaborative filtering network $F$, and use RMSprop to learn the discriminators $D_{\text{u}}$ and $D_{\text{v}}$, where learning rate is set to 0.0001. For $D_{\text{u}}$ and $D_{\text{v}}$, we set the number of hidden layers to 2 and the numbers of hidden nodes are respectively 50 and 100. For $F$, we set three hidden layers each of which consists of 32 hidden nodes. We use standard Gaussian distribution as the prior of the embeddings.

\subsection{Experimental Analysis}
\subsubsection{Recommendation Performance (RQ1)}

Tables \ref{Tab_ML100k_TOPN}, \ref{Tab_ML1M_TOPN}, \ref{Tab_YELP_TOPN}, \ref{Tab_DM_TOPN} and \ref{Tab_Pinterest_TOPN} show the results of top-$k$ recommendation on the five datasets, respectively, where $k = \{5, 10, 20\}$. 

\begin{table}[t]
	\caption{Recommendation Performance on MovieLens 100K. The best runs per metric are marked in boldface. The best runs per metric among robust recommendation methods CDAE, APR, and ACAE are underlined.}
	\begin{center}
		\setlength{\tabcolsep}{3pt}{
			\begin{tabular}{l|c|c|c|c|c|c}
				\hline 
				\multirow{2}*\textbf{ }
				\rule{0pt}{10pt}
				&\multicolumn{6}{c}{MovieLens 100K} \\
				\cline{2-7}
			    \rule{0pt}{10pt}
			    &HR &HR &HR &NDCG &NDCG &NDCG\\ 
			    &@5 &@10 &@20 &@5 &@10 &@20\\ \hline
				\rule{0pt}{10pt}CDAE 
				&0.4284&0.6331&0.7996&0.2855&0.3511&0.3934\\
				\hline
				\rule{0pt}{11pt}APR 
				&\underline{0.4772}&\underline{0.6755}&\underline{0.8261}&\underline{0.3253}&\underline{0.3896}&\underline{0.4276} \\ \hline
				\rule{0pt}{11pt}ACAE 
				&0.4602&0.6437&0.8049&0.3107&0.3697&0.4106 \\ \hline
				\rule{0pt}{11pt}CFGAN &0.2810&0.4422&0.632&0.1921&0.2438&0.2913\\ \hline
				\rule{0pt}{11pt}NeuMF 
				&0.4645&0.6257&0.7943&0.3183&0.3704&0.4128\\ \hline
				\rule{0pt}{11pt}AVB
			    &0.3648&0.5514&0.7328&0.2402&0.3000 &0.3620\\ \hline
				\rule{0pt}{11pt}VAEGAN
				&0.3107&0.4634&0.6459&0.2044&0.2531&0.2992\\ \hline
				\rule{0pt}{11pt}CVAE-GAN
				&0.2609&0.4008&0.5832&0.1673&0.2117&0.2571\\ \hline
				\rule{0pt}{11pt}RecVAE
				&0.4793&0.6448&0.8028&0.3216&0.3753&0.4157\\ \hline
				\rule{0pt}{11pt}DAVE-adv &0.4634&0.6288&0.7794&0.3193&0.3727&0.4108\\ \hline
				\rule{0pt}{11pt}DAVE+aae &0.4942&\textbf{0.6776}&0.8282&0.3317&0.3907&0.4291\\ \hline
				\rule{0pt}{11pt}DAVE&\textbf{0.4995}&0.6723&\textbf{0.8293}&\textbf{0.3415}&\textbf{0.3971}&\textbf{0.4369}\\ \hline
		\end{tabular}}
	\end{center}
	\label{Tab_ML100k_TOPN}
\end{table}

\begin{table}[t]  
	\caption{Recommendation Performance on MovieLens 1M. The best runs per metric are marked in boldface. The best runs per metric among robust recommendation methods CDAE, APR, and ACAE are underlined.}
	\begin{center}
		\setlength{\tabcolsep}{3pt}{
			\begin{tabular}{l|c|c|c|c|c|c}
				\hline 
				\multirow{2}*\textbf{ }
				\rule{0pt}{10pt}
				&\multicolumn{6}{c}{MovieLens 1M} \\
				\cline{2-7}
				\rule{0pt}{10pt}
			    &HR &HR &HR &NDCG &NDCG &NDCG\\ 
			    &@5 &@10 &@20 &@5 &@10 &@20\\ \hline
				\rule{0pt}{11pt}CDAE &0.4343&0.6134&0.7882&0.2948&0.3527&0.3970 \\ \hline
				\rule{0pt}{11pt}APR &0.4603&0.6396&\underline{0.8167}&0.3148&0.3728&0.4176 \\ \hline
				\rule{0pt}{11pt}ACAE &\underline{0.5002}&\underline{0.6649}&0.8164&\underline{0.3473}&\underline{0.4004}&\underline{0.4388}\\ \hline
				\rule{0pt}{11pt}CFGAN &0.3070&0.4594&0.6339&0.2077&0.2568&0.3007\\ \hline
				\rule{0pt}{11pt}NeuMF &0.5089&0.6833&0.8321&0.3562&0.4124&0.4503\\ \hline
				\rule{0pt}{11pt}AVB &0.3891&0.5705&0.7500&0.2582&0.3167&0.4576\\ \hline
				\rule{0pt}{11pt}VAEGAN
				&0.3166&0.4652&0.6512&0.2118&0.2596&0.3064\\ \hline
				\rule{0pt}{11pt}CVAE-GAN
				&0.2877&0.4348&0.6308&0.1879&0.2365&0.2847\\ \hline
				\rule{0pt}{11pt}RecVAE &0.5371&0.6993&0.8467&0.3729&0.4257&0.4631  \\ \hline
				\rule{0pt}{11pt}DAVE-adv &0.4752&0.6575&0.8270&0.3242&0.3832&0.4261  \\ \hline
				\rule{0pt}{11pt}DAVE+aae &0.5248&0.6909&0.8397&0.3625&0.4162&0.4541\\ \hline
				\rule{0pt}{11pt}DAVE&\textbf{0.5417}&\textbf{0.7185}&\textbf{0.8518}&\textbf{0.3761}&\textbf{0.4334}&\textbf{0.4671}\\ \hline
		\end{tabular}}
	\end{center}
	\label{Tab_ML1M_TOPN}
\end{table}

\begin{table}[t]
	\caption{Recommendation Performance on Yelp. The best runs per metric are marked in boldface. The best runs per metric among robust recommendation methods CDAE, APR, and ACAE are underlined.}
	\begin{center}
		\setlength{\tabcolsep}{3pt}{
			\begin{tabular}{l|c|c|c|c|c|c}
				\hline 
				\multirow{2}*\textbf{ }
				\rule{0pt}{10pt}
				&\multicolumn{6}{c}{Yelp} \\
				\cline{2-7}
			    \rule{0pt}{10pt}
		    	&HR &HR &HR &NDCG &NDCG &NDCG\\ 
		    	&@5 &@10 &@20 &@5 &@10 &@20\\ \hline
				\rule{0pt}{11pt}CDAE 
				&0.3231&0.4444&0.5963&0.2289&0.2680&0.3064 \\ 
				\hline
				\rule{0pt}{11pt}APR &\underline{0.6494}&\underline{0.7920}&\underline{\textbf{0.9048}}&\underline{0.4810}&\underline{0.5274}&\underline{0.5560} \\ \hline
				\rule{0pt}{11pt}ACAE &0.6125&0.7569&0.8746&0.4527&0.4996&0.5294 \\ \hline
				\rule{0pt}{11pt}CFGAN &0.3027&0.4252&0.5626&0.2110&0.2504&0.2850 \\ \hline
				\rule{0pt}{11pt}NeuMF &0.6529&0.7838&0.8793&0.4836&0.5262 &0.5505\\ \hline
				\rule{0pt}{11pt}AVB &0.3244&0.4476&0.5944&0.2297&0.2694 &0.3064\\ \hline
				\rule{0pt}{11pt}VAEGAN
				&0.3273&0.4506&0.6044&0.2320&0.2717&0.3105\\ \hline
				\rule{0pt}{11pt}CVAE-GAN
				&0.3227&0.4495&0.5976&0.2293&0.2701&0.3074\\ \hline
				\rule{0pt}{11pt}RecVAE &0.6464&0.7843&0.8936&0.4866&0.5313&0.5590  \\ \hline
				\rule{0pt}{11pt}DAVE-adv &0.6192&0.7643&0.8744&0.4504&0.4976&0.5256  \\ \hline
				\rule{0pt}{11pt}DAVE+aae &0.6687&0.8022&0.9025&0.4980&0.5415&0.5670 \\	\hline
				\rule{0pt}{11pt}DAVE&\textbf{0.6688}&\textbf{0.8032}&0.9015&\textbf{0.5018}&\textbf{0.5456}&\textbf{0.5706}\\ \hline
		\end{tabular}}
	\end{center}
	\label{Tab_YELP_TOPN}
\end{table}

\begin{table}[t]
	\caption{Recommendation Performance on Digital Music. The best runs per metric are marked in boldface. The best runs per metric among robust recommendation methods CDAE, APR, and ACAE are underlined.}
	\begin{center}
		\setlength{\tabcolsep}{3pt}{
			\begin{tabular}{l|c|c|c|c|c|c}
				\hline 
				\multirow{2}*\textbf{ }
				\rule{0pt}{10pt}
				&\multicolumn{6}{c}{Digital Music} \\
				\cline{2-7}
			    \rule{0pt}{10pt}
			    &HR &HR &HR &NDCG &NDCG &NDCG\\ 
		     	&@5 &@10 &@20 &@5 &@10 &@20\\ \hline
				\rule{0pt}{11pt}CDAE 
				&0.2018&0.3001&0.4215&0.1375&0.1692&0.1997 \\ 
				\hline
				\rule{0pt}{11pt}APR &\underline{0.4861}&\underline{0.6145}&\underline{0.7435}&\underline{0.3500}&\underline{0.3908}&\underline{0.4235} \\ \hline
				\rule{0pt}{11pt}ACAE &0.4562&0.5930&0.7319&0.3355&0.3796&0.4147 \\ \hline
				\rule{0pt}{11pt}CFGAN &0.1978&0.2896&0.4066&0.1340&0.1636&0.1931 \\ \hline
				\rule{0pt}{11pt}NeuMF &0.3534&0.4707&0.6049&0.2597&0.2974 &0.3312\\ \hline
				\rule{0pt}{11pt}AVB &0.2000&0.2956&0.4164&0.1367&0.1675 &0.1980\\ \hline
				\rule{0pt}{11pt}VAEGAN
				&0.2081&0.3059&0.4348&0.1416&0.1730&0.2054\\ \hline
				\rule{0pt}{11pt}CVAE-GAN
				&0.2058&0.3048&0.4284&0.1402&0.1721&0.2032\\ \hline
				\rule{0pt}{11pt}RecVAE
				&0.4128&0.5226&0.6504&0.3190&0.3544&0.3866\\ \hline
				\rule{0pt}{11pt}DAVE-adv &0.4328&0.5680&0.7156&0.3131&0.3566&0.3939  \\ \hline
				\rule{0pt}{11pt}DAVE+aae &0.4760&0.6192&0.7594&0.3420&0.3883&0.4239 \\	\hline
				\rule{0pt}{11pt}DAVE&\textbf{0.4872}&\textbf{0.6269}&\textbf{0.7651}&\textbf{0.3555}&\textbf{0.4007}&\textbf{0.4357}\\ \hline
		\end{tabular}}
	\end{center}
	\label{Tab_DM_TOPN}
\end{table}

At first, from Tables \ref{Tab_ML100k_TOPN}, \ref{Tab_ML1M_TOPN}, \ref{Tab_YELP_TOPN}, \ref{Tab_DM_TOPN} and \ref{Tab_Pinterest_TOPN}, we can see that on MovieLens 100K, MovieLens 1M, Yelp and Digital Music, DAVE shows better performance than the robust recommendation methods CDAE, APR and ACAE with respect to HR@$k$, except for HR@10 on MovieLens 100K and HR@20 on Yelp. 

We can also note that DAVE consistently outperforms CDAE, APR and ACAE with respect to NDCG@$k$ on five datasets. In particular, on MovieLens 100K, compared with the most competitive method APR, DAVE increases the NDCG@5 by 5\%, NDCG@10 by 1.9\%, and NDCG@20 by 2.2\% (see Table \ref{Tab_ML100k_TOPN}); on MovieLens 1M, compared to the most competitive method ACAE, DAVE increases the NDCG@5 by 8.3\%, NDCG@10 by 8.2\%, and NDCG@20 by 6.4\% (see Table \ref{Tab_ML1M_TOPN}); on Yelp, compared to the best competitor APR, DAVE increases the NDCG@5 by 4.3\%, NDCG@10 by 3.5\%, and NDCG@20 by 2.6\% (see Table \ref{Tab_YELP_TOPN}); on Digital Music, compared to the best competitor APR, DAVE increases the NDCG@5 by 1.5\%, NDCG@10 by 1.5\%, and NDCG@20 by 2.8\% (see Table \ref{Tab_DM_TOPN}); on Pinterest, compared to the best competitor APR, DAVE increases the NDCG@5 by 1.3\%, NDCG@10 by 0.9\%, and NDCG@20 by 1\% (see Table \ref{Tab_Pinterest_TOPN}). We argue that these improvements are mainly due to the better expressiveness of DAVE. Unlike the existing robust recommendation methods that assume user preference distribution is a single modal, DAVE is able to handle the multi-modality of user-item interaction data so that true user preference distributed around different modes can be captured. At last, we also note the exception that DAVE is slightly inferior to APR with respect to HR@$k$ on Pinterest. This is partly because in Pinterest the number of users is far more than the number of items. Such unbalance reduces the preference diversity revealed by the data and consequently hinders DAVE from best capturing the multi-modality of the preference distributions.

We can also see that DAVE outperforms NeuMF and CFGAN on all datasets. For NeuMF, this is because that DAVE can model different noise distributions via VAE during the representation learning for different users and items, which leads to more robust embeddings than naive neural collaborative filtering. Note that CFGAN also combines GAN as well as adversarial training with collaborative filtering, but it directly generates user-item interaction vectors for collaborative filtering rather than learns latent representations for users and items, which is in contrast with DAVE. Although the output of a certain hidden layer of the generative model of CFGAN can serve as user latent representation, learning only the user latent representation is not enough to model the non-linear interactions between users and items, so it is difficult for CFGAN to effectively capture the preference of user to item, which leads CFGAN to almost the worst performance. 

Finally, we can observe that DAVE outperforms AVB, VAEGAN, CVAE-GAN and RecVAE on all datasets. Although AVB and VAEGAN also focus on tackling the single-modality problem of VAE by utilizing GAN as well as adversarial training, they cannot provide  personalized noise reduction for different users. RecVAE improves VAE with a new architecture for the encoder, but it cannot capture multimodal preference distributions. In addition, AVB, VAEGAN, CVAE-GAN and RecVAE learn latent representations only for users, which are not enough to model the non-linear interactions between users and items. DAVE infers distributions over embeddings for both users and items by two VAEs, which combines the advantages of the user-based methods and item-based methods for collaborative filtering.
\begin{table}[t]
	\caption{Recommendation Performance on Pinterest. The best runs per metric are marked in boldface. The best runs per metric among robust recommendation methods CDAE, APR, and ACAE are underlined.}
	\begin{center}
	
		\setlength{\tabcolsep}{2pt}{
		
			\begin{tabular}{l|c|c|c|c|c|c}
				
				\hline 
				\multirow{2}*\textbf{ }
				\rule{0pt}{10pt}
				&\multicolumn{6}{c}{Pinterest} \\
				\cline{2-7}
				\rule{0pt}{10pt}
				&HR &HR &HR &NDCG &NDCG &NDCG\\ 
				&@5 &@10 &@20 &@5 &@10 &@20\\ \hline
				\rule{0pt}{11pt}CDAE 
				&0.3303&0.4814&0.6395&0.2174&0.2662&0.3063 \\ 
				\hline
				\rule{0pt}{11pt}APR	  &\underline{\textbf{0.7246}}&\underline{\textbf{0.8884}}&\underline{\textbf{0.9704}}&\underline{0.5157}&\underline{0.5691}&\underline{0.5902} \\ \hline
				\rule{0pt}{11pt}ACAE &0.7086&0.8756&0.9663&0.5024&0.5569&0.5802 \\ \hline
				\rule{0pt}{11pt}CFGAN &0.1596&0.2628&0.4118&0.1016&0.1348&0.1722 \\ \hline
				\rule{0pt}{11pt}NeuMF &0.6890&0.8664&0.9619&0.4831&0.5410 &0.5656\\ \hline
				\rule{0pt}{11pt}AVB &0.2731&0.4611&0.6452&0.1407&0.2013 &0.2480\\ \hline
				\rule{0pt}{11pt}VAEGAN
				&0.1778&0.2901&0.4576&0.1134&0.1494&0.1915\\ \hline
				\rule{0pt}{11pt}CVAE-GAN
				&0.1739&0.2863&0.4530&0.1108&0.1469&0.1887\\ \hline
				\rule{0pt}{11pt}RecVAE &0.6851&0.8371&0.9341&0.5030&0.5525&0.5773  \\ \hline
				\rule{0pt}{11pt}DAVE-adv &0.7020&0.8698&0.9617&0.5016&0.5563&0.5800  \\ \hline
				\rule{0pt}{11pt}DAVE+aae &0.6864&0.8576&0.9562&0.4892&0.5450&0.5704 \\	\hline
				\rule{0pt}{11pt}DAVE&0.7219&0.8798&0.9663&\textbf{0.5226}&\textbf{0.5741}&\textbf{0.5963}\\ \hline
		\end{tabular}}
	\end{center}
	\label{Tab_Pinterest_TOPN}
\end{table}

\begin{figure*}
	\centering
	\begin{minipage}[c]{0.24\textwidth}
		\centering
		\subfigure[HR@5]{\includegraphics[width=1\textwidth]{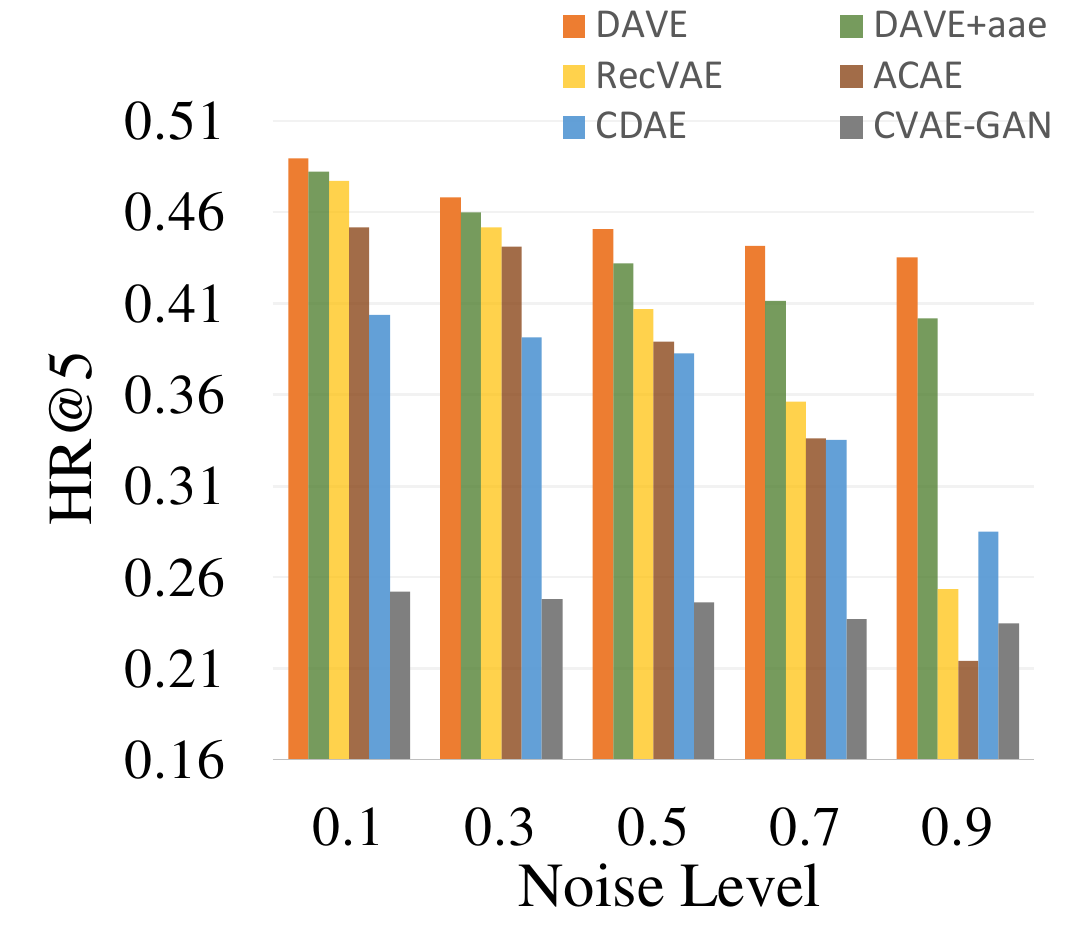}}
	\end{minipage}
	\begin{minipage}[c]{0.24\textwidth}
		\centering
		\subfigure[NDCG@5]{\includegraphics[width=1\textwidth]{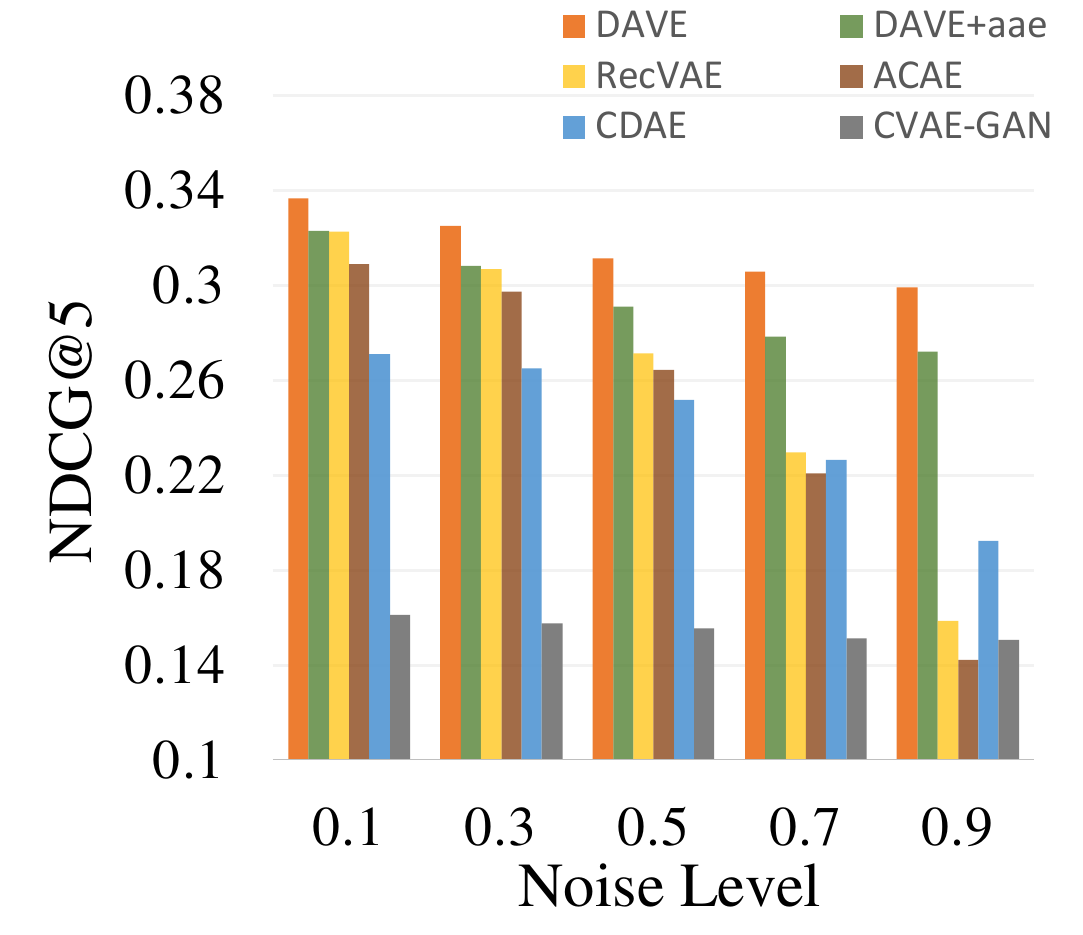}}
	\end{minipage}	
	\begin{minipage}[c]{0.24\textwidth}
		\centering
		\subfigure[HR@10]{\includegraphics[width=1\textwidth]{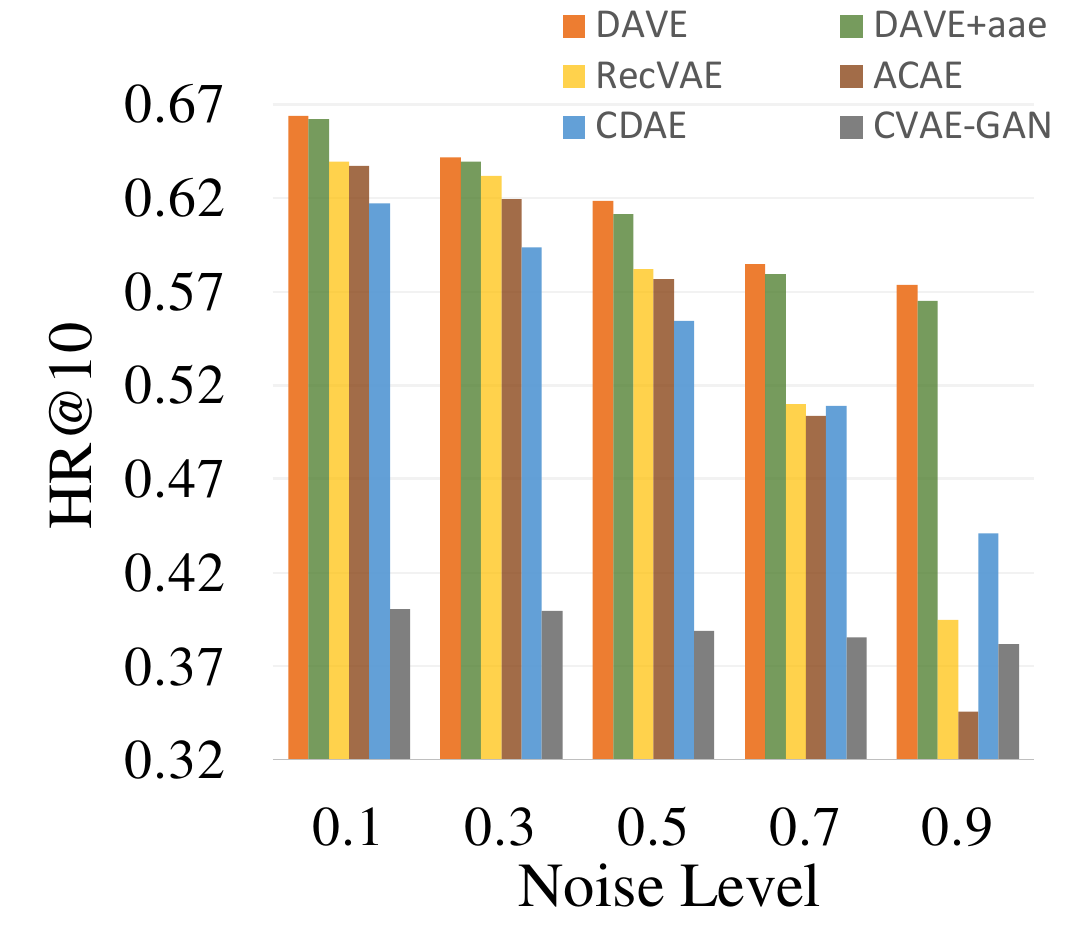}}
	\end{minipage}
	\begin{minipage}[c]{0.24\textwidth}
		\centering
		\subfigure[NDCG@10]{\includegraphics[width=1\textwidth]{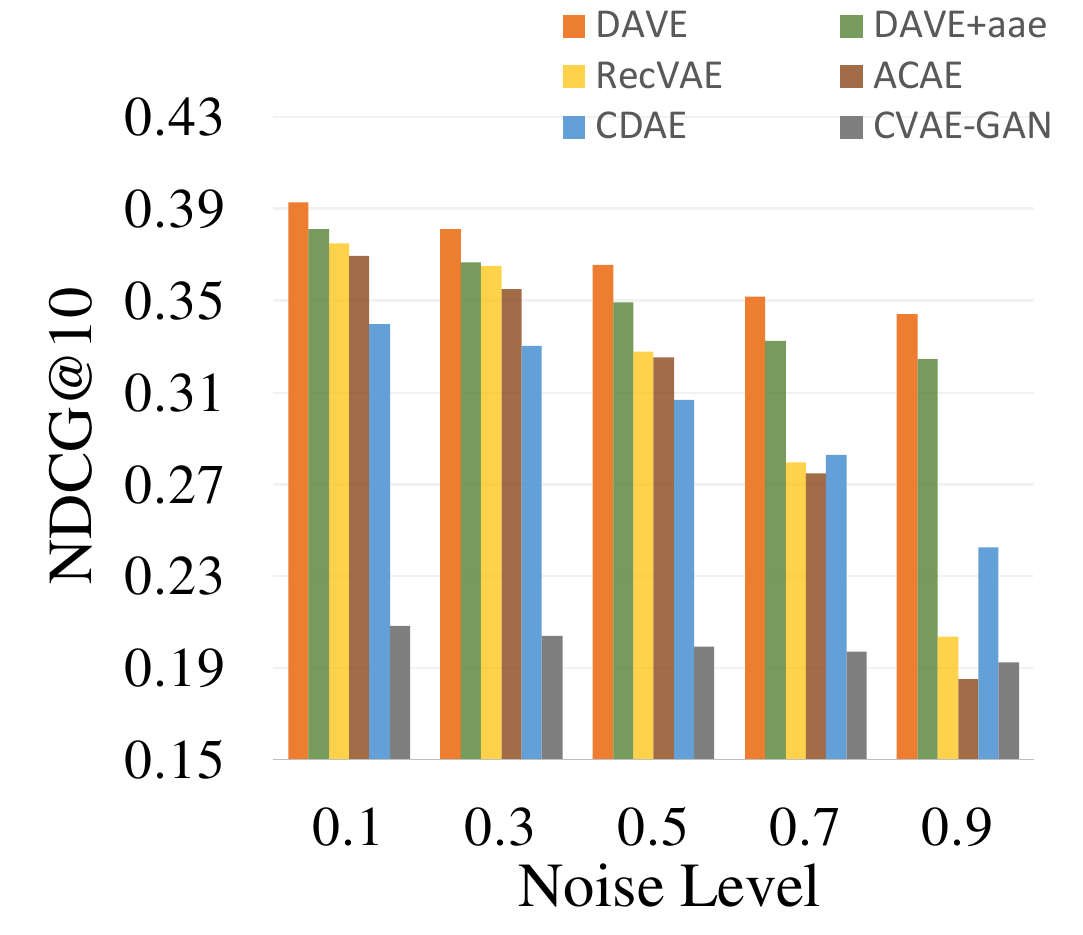}}
	\end{minipage}
	\caption{Noise tolerability on MovieLens 100K.}
	\label{robust_HR}
\end{figure*}

\subsubsection{Noise Tolerability (RQ2)}

Now we investigate the robustness of DAVE by comparing it with its variant DAVE+aae over MovieLens 100K. We also compare DAVE with RecVAE, ACAE, CDAE, and CAVE-GAN since they can address noise data too. Here we want to verify DAVE has better recommendation performance as well as better noise tolerability in the face of noisy interaction. For each user $u$ and item $v$ in the testing set, we intentionally inject some noise to the data through two steps: first randomly choose a noise level (i.e., the ratio of the noisy interactions) and then randomly flipping over some entries of their interaction vectors $\boldsymbol{u}$ and $\boldsymbol{v}$ with respect to the chosen noise level, by which we simulate the scenario that different users or items have different noise levels. Figure \ref{robust_HR} reports the results in terms of HR@$k$ and NDCG@$k$ with $k = 5,10$ at the noise levels 0.1, 0.3, 0.5, 0.7, and 0.9. 

From Figure \ref{robust_HR} we can observe that the performance of all methods degrades as noise level increases. At different noise levels, however, DAVE consistently exhibits better performance than all the alternative methods, and the greater the noise level (i.e., the more the noise added), the bigger the gap between DAVE and the alternative methods. DAVE+aae generates the user or item embeddings with a point estimate produced by Adversarial Auto-encoder (AAE) \cite{Makhzani2016Adversarial}. In contrast, DAVE generates the embedding for a user or an item by sampling from an inferred embedding distribution unique to that user or item. The result shows that inferring unique embedding distribution for different users and items brings DAVE the better noise tolerability. Similarly, DAVE shows much better noise tolerability than RecVAE, ACAE, CDAE, and CAVE-GAN as it can generate more expressive preference embeddings due to its adaptability to the different noise distributions and the ability to capture the multi-modality of the preference distributions.

\subsubsection{Model Expressiveness (RQ3)}

In the experiments, the posterior distributions of embeddings unique to different users and items are Gaussian. For a user $u$, the variational inference network $Q_{\text{u}}$ in UserAVE will generate a pair of mean and standard deviation, $(\boldsymbol{\mu}_u, \boldsymbol{\sigma}_u)$, which defines the posterior distributions of embeddings of that user. Similarly, for an item $v$, the variational inference network $Q_{\text{v}}$ in ItemAVE will generate the pair $(\boldsymbol{\mu}_v, \boldsymbol{\sigma}_v)$ to define the posterior distributions of embeddings of that item. To evaluate the expressiveness of DAVE, we will check the distributions of $(\boldsymbol{\mu}_u, \boldsymbol{\sigma}_u)$ and $(\boldsymbol{\mu}_v, \boldsymbol{\sigma}_v)$ inferred by DAVE and DAVE-adv on MovieLens 100K. Particularly, for each pair $(\boldsymbol{\mu}_u, \boldsymbol{\sigma}_u)$, we concatenate $\boldsymbol{\mu}_u$ and $\boldsymbol{\sigma}_u$ to form a new vector to represent the posterior distribution defined by $(\boldsymbol{\mu}_u, \boldsymbol{\sigma}_u)$, and then visualize the distribution of these new vectors in a 2-dimensional space using t-SNE algorithm \cite{Maaten2008Viualizing}. The same process is also applied to each $(\boldsymbol{\mu}_v, \boldsymbol{\sigma}_v)$. 

\begin{figure}[t]
	\begin{minipage}[t]{0.23\textwidth}
		\centering
		\subfigure[DAVE]{\includegraphics[width=1.1\textwidth]{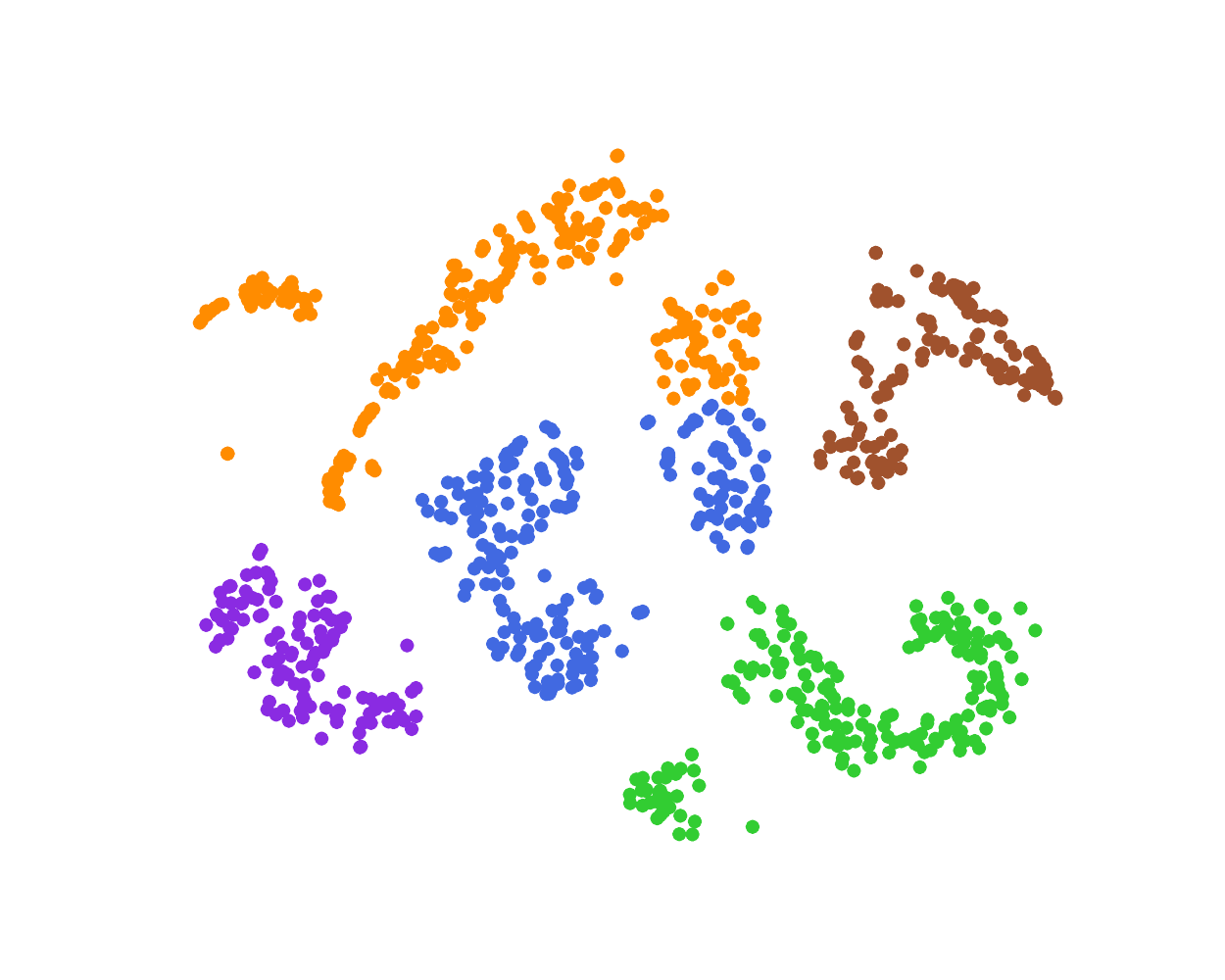}}
	\end{minipage}
	\begin{minipage}[t]{0.23\textwidth}
		\centering
		\subfigure[DAVE-adv]{\includegraphics[width=1.1\textwidth]{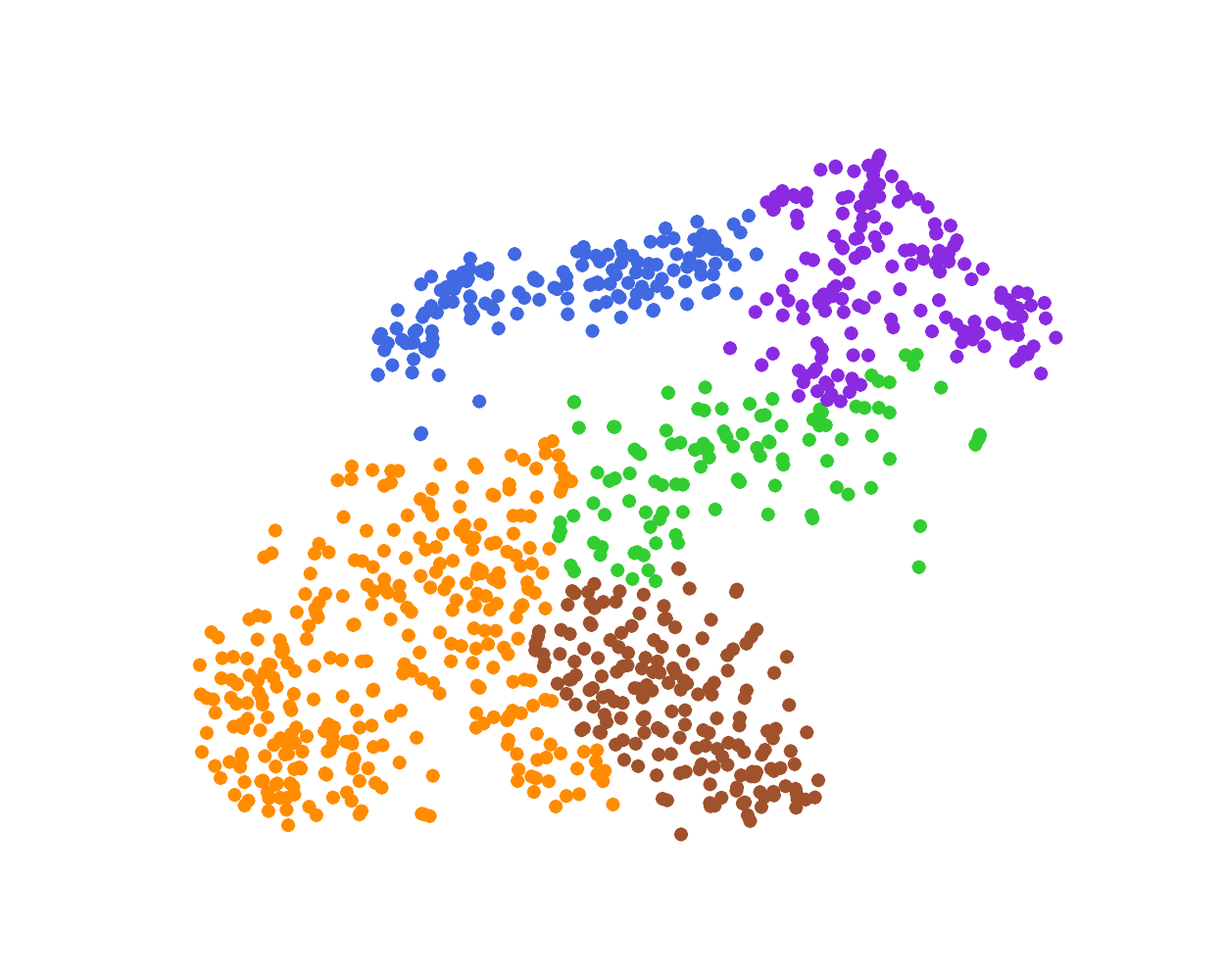}}
	\end{minipage}
	\caption{Visualization of the distribution of the posterior distributions of user embeddings learned by (a) DAVE and (b) DAVE-adv in MovieLens 100K.}
	\label{DAVE-adv_User}
\end{figure}

\begin{figure}[t]
	\begin{minipage}[t]{0.23\textwidth}
		\centering
		\subfigure[DAVE]{\includegraphics[width=1.1\textwidth]{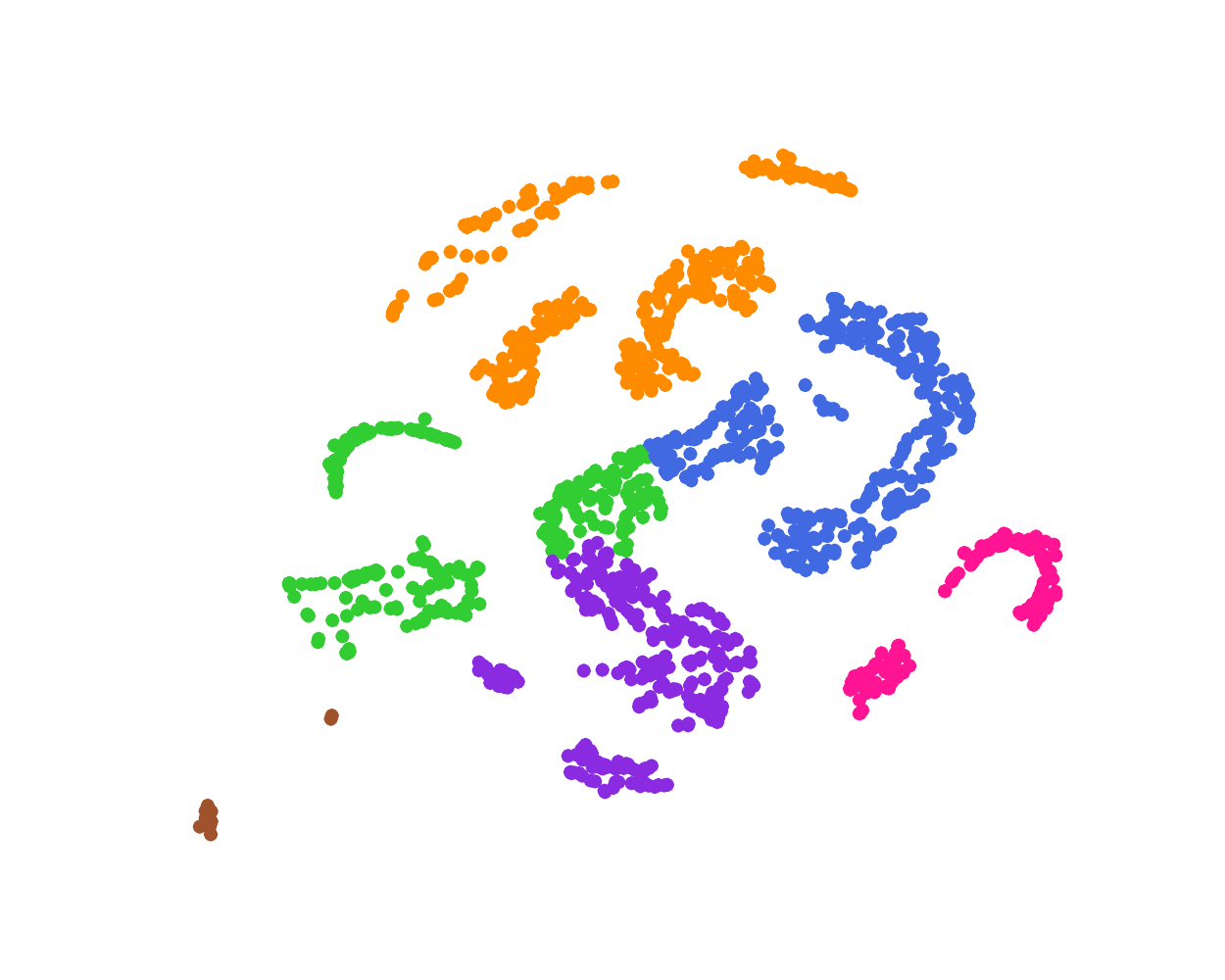}}
	\end{minipage}
	\begin{minipage}[t]{0.23\textwidth}
		\centering
		\subfigure[DAVE-adv]{\includegraphics[width=1.1\textwidth]{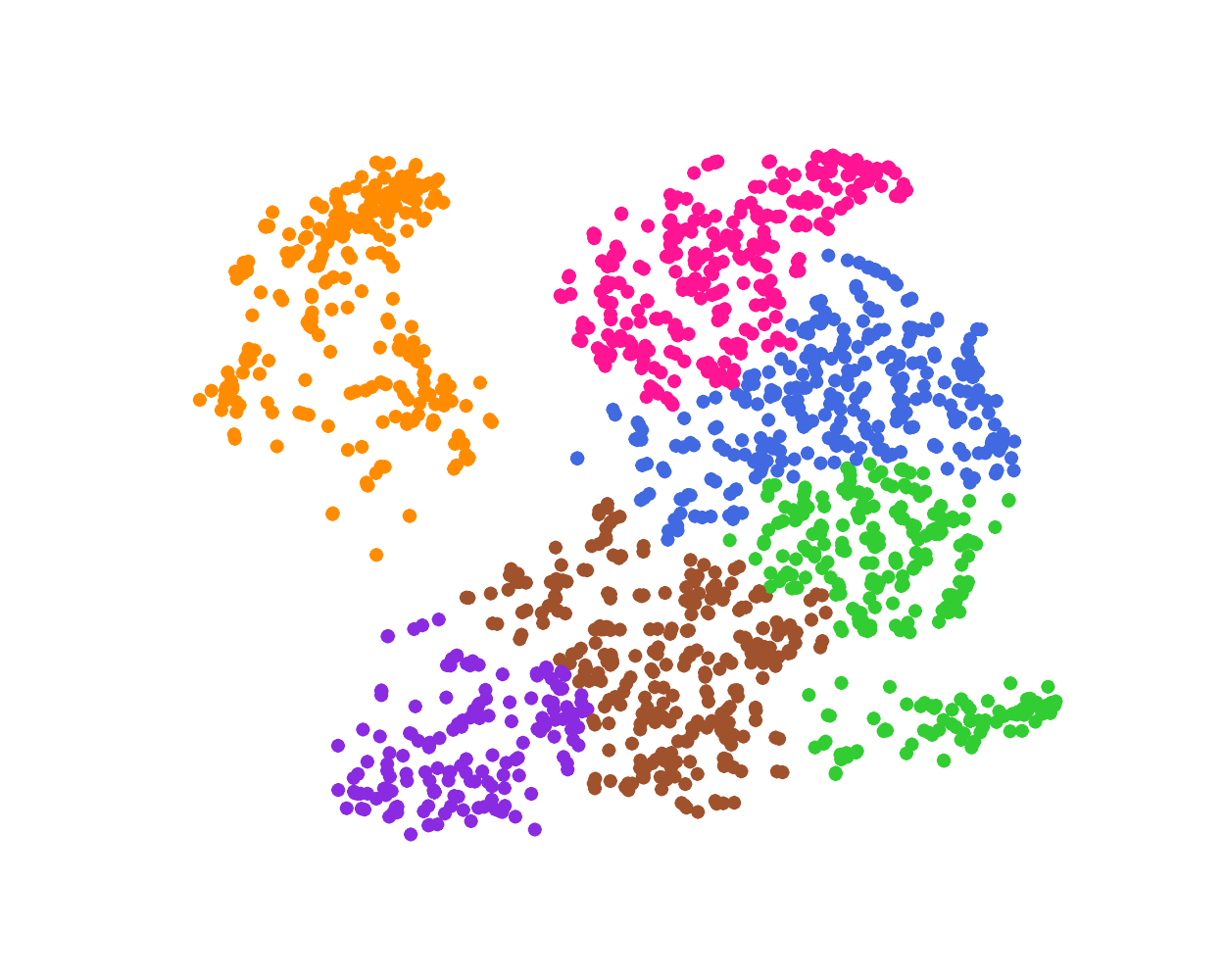}}
	\end{minipage}
	\caption{Visualization of the distribution of the posterior distributions of item embeddings learned by (a) DAVE and (b) DAVE-adv in MovieLens 100K.}
	\label{DAVE-adv_Item}
\end{figure}

Figures \ref{DAVE-adv_User} and \ref{DAVE-adv_Item} show the visualization of the posterior embedding distributions of 943 users and 1682 items in MovieLens 100K, respectively, where a point represents the 2-dimensional projection of a concatenating vector and the points belonging to the same cluster are of the same color.

We can see that the points representing the posterior embedding distributions inferred by DAVE are obviously separated into multiple clusters (distributions) shown in Figures \ref{DAVE-adv_User}(a) and \ref{DAVE-adv_Item}(a), and the gaps between the clusters inferred by DAVE are far more significant than the gaps between the clusters inferred by DAVE-adv shown in Figures \ref{DAVE-adv_User}(b) and \ref{DAVE-adv_Item}(b). We argue that such difference is caused by the different effects offered by KL-divergence and adversarial training. In DAVE-adv, in order to compute the KL-divergence, the posterior distribution is explicitly represented with a Gaussian, and minimizing the KL-divergence encourages the posterior distributions to be close to a common prior distribution, which consequently makes the posterior distributions inferred by DAVE-adv tend to be single modal and unexpressive. On the contrary, although the adversarial training in DAVE approximates the minimizing of KL-divergence, it offers the flexibility without the requirement to explicitly represent the posterior distribution. Such flexibility makes it possible for DAVE to infer complex posterior distributions that are multimodal and cannot to be explicitly formulated, which improves the expressiveness of the user embeddings drawn from the inferred posterior distributions and benefits the capturing of the diversity of user preference. 

\begin{figure}[t]
	\begin{minipage}[t]{0.23\textwidth}
		\centering
		\subfigure[HR@10]{\includegraphics[width=1\textwidth]{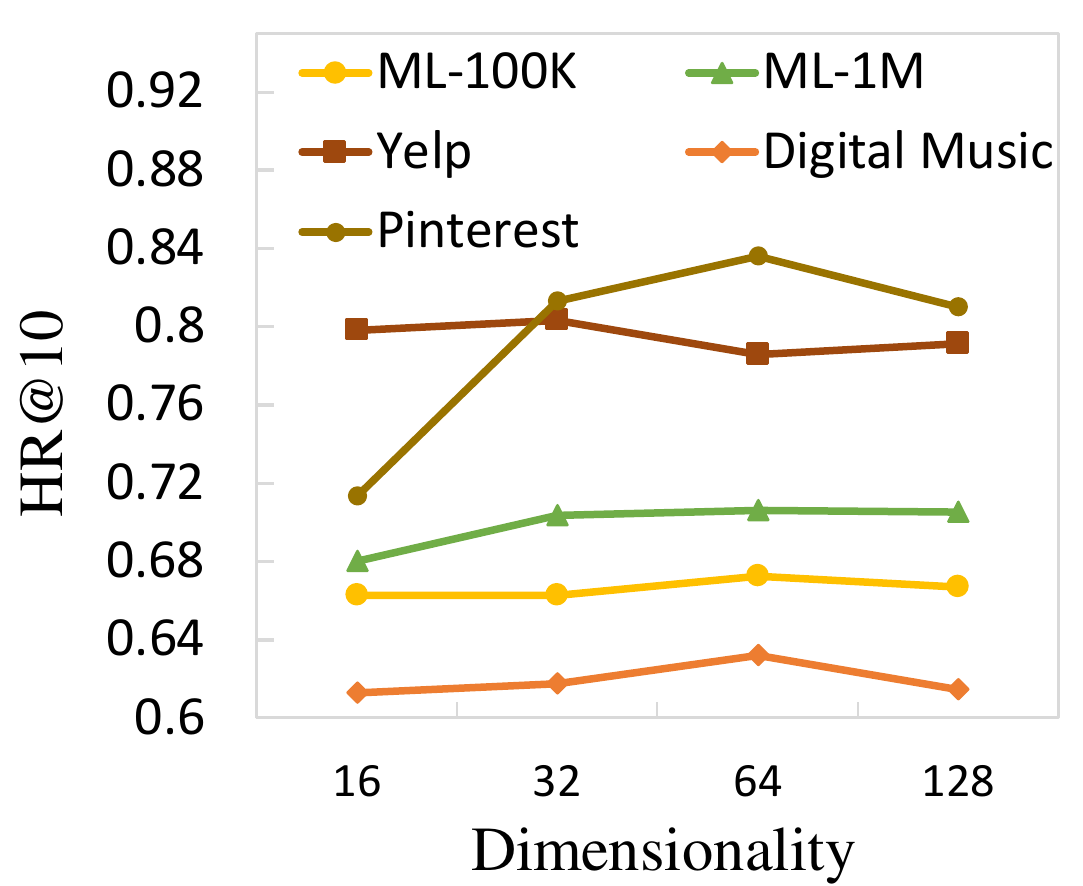}}
	\end{minipage}
	\begin{minipage}[t]{0.23\textwidth}
		\centering
		\subfigure[NDCG@10]{\includegraphics[width=1\textwidth]{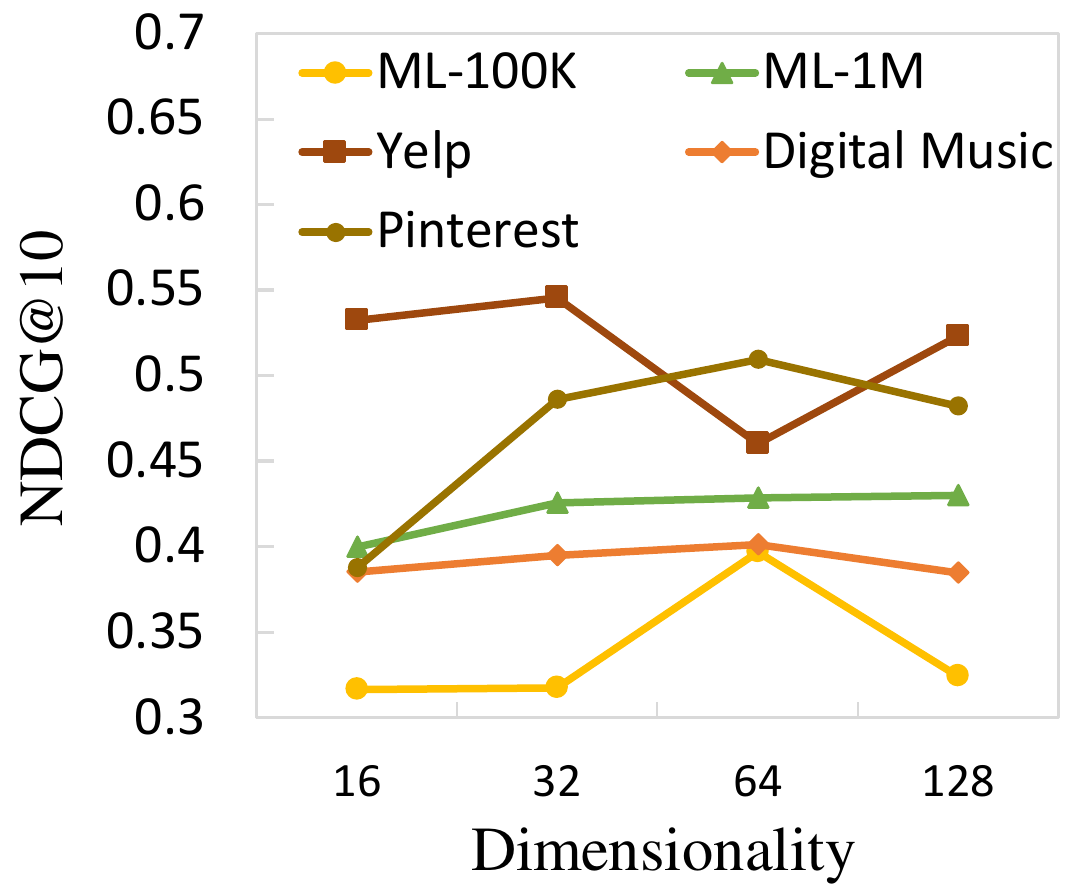}}
	\end{minipage}
	\caption{Tuning of embedding dimensionality.}
	\label{Dim@10}
\end{figure}

\subsubsection{Tuning of Hyper-parameters (RQ4)}
\begin{figure}[!t]
	\begin{minipage}[t]{0.23\textwidth}
		\centering
		\subfigure[HR@10]{\includegraphics[width=1\textwidth]{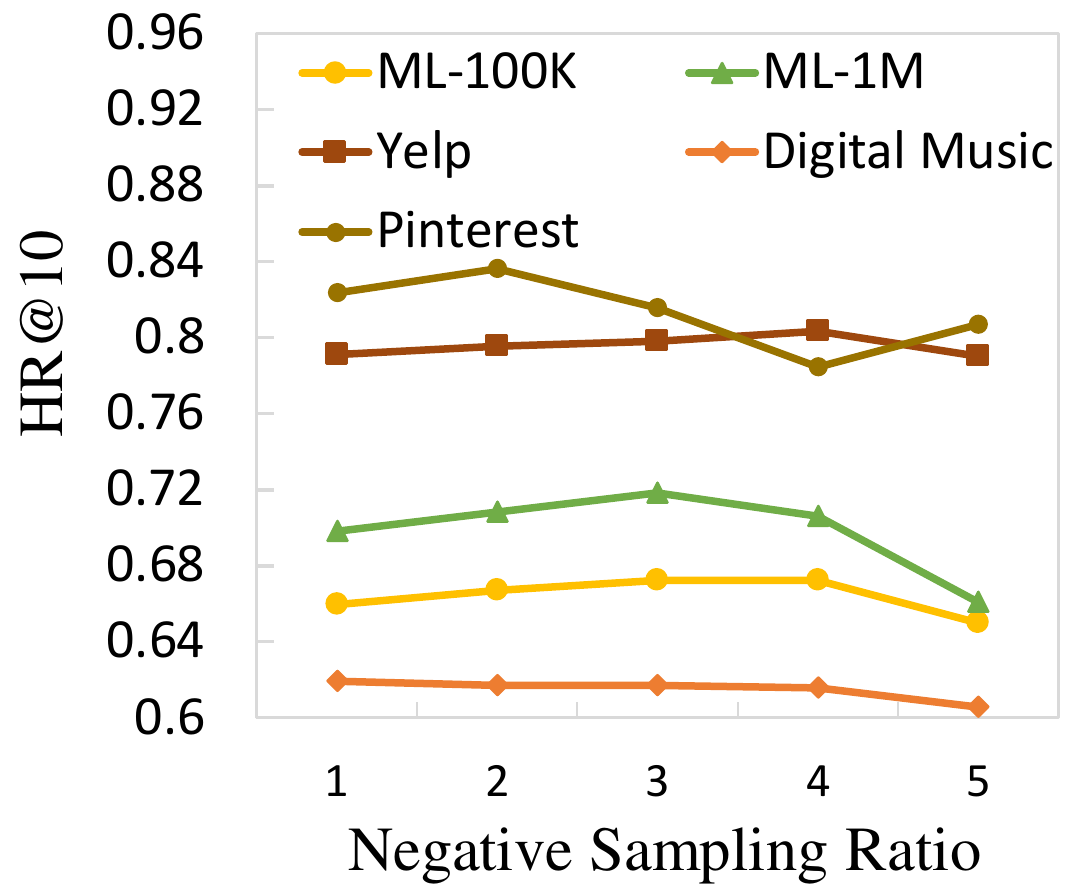}}
	\end{minipage}
	\begin{minipage}[t]{0.23\textwidth}
		\centering
		\subfigure[NDCG@10]{\includegraphics[width=1\textwidth]{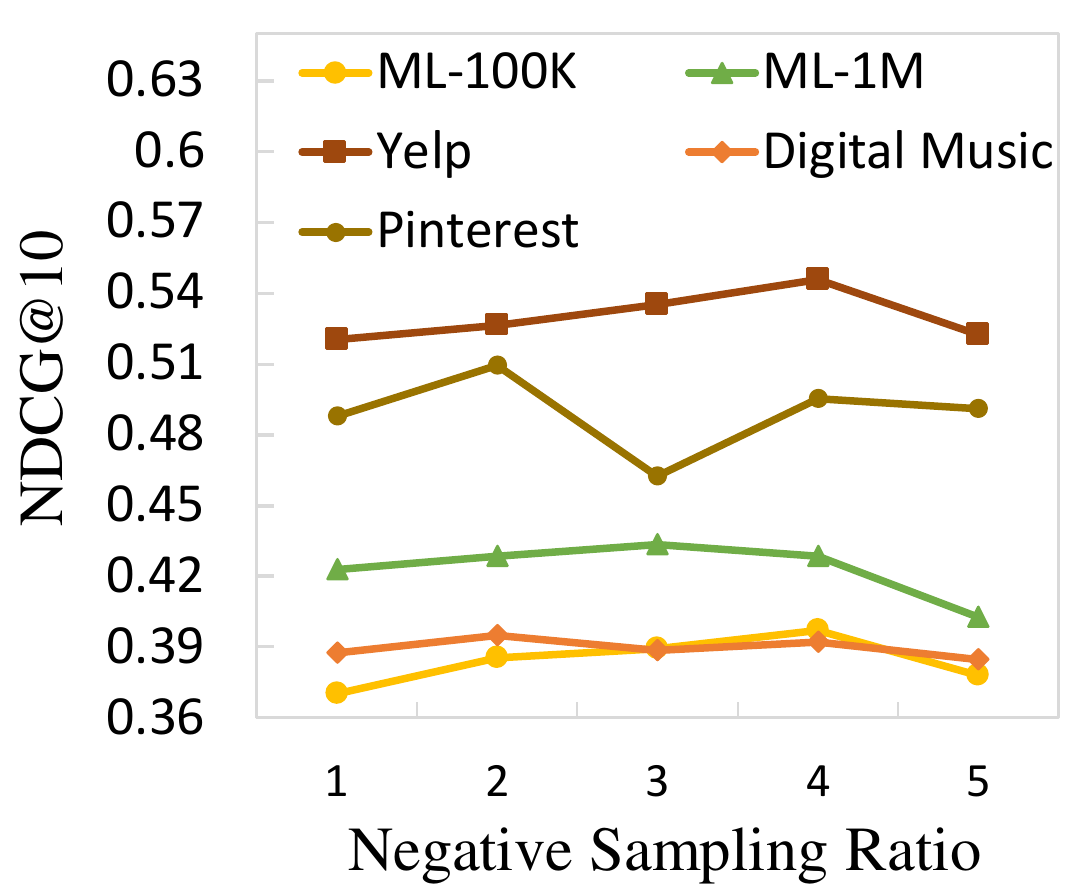}}
	\end{minipage}
	\caption{Tuning of negative sampling ratio.}
	\label{Neg@10}
\end{figure}

Now we tune two hyper-parameters of DAVE, embedding dimensionality and negative sampling ratio, in terms of HR@10 and NDCG@10 over five validation sets. The results are shown in Figures \ref{Dim@10} and \ref{Neg@10}, respectively. 

From Figure 5, we can see that on MovieLens 100K, MovieLens 1M, Digital Music and Pinterest, the optimal embedding dimensionality is 64, while on Yelp it is 32. Basically the performance of DAVE improves first with the increase of embedding dimensionality, then degrades due to overfitting incurred by excessive embedding dimensionality. We also note that on Yelp, after the optimal embedding dimensionality 32, the performance of DAVE is on downward trend with some fluctuates that might be incurred by random initialization of parameters. 

Figure 6 shows the effect of negative sampling ratio on the performance of DAVE. We can see that with the increase of negative sampling ratio from 1 to 5, the HR@10 and NDCG@10 grow first, then drop. On Yelp and MovieLens 100K, DAVE achieves the best performance at the negative sampling ratio of 4, and on Digital Music and Pinterest, DAVE achieves the best performance at the negative sampling ratio of 2 while on MovieLens 1M, DAVE achieves the best performance at the negative sampling ratio of 3. The observation also implies that excessively high negative sampling ratio may mistakenly lead to more false-negative samples, which results in reduced robustness and weaker generalization performance of DAVE.

\section{RELATED WORK}
In this section, we briefly review related work on the traditional recommender systems and the robust recommendation.

\subsection {Traditional Recommender Systems}
In traditional recommender systems, collaborative filtering is the most widely used technique for personalized recommendation, which aims to predict user preference from historical user-item interactions, with a learnable interaction function of informative representations of users and items that capture the collaborative signals, i.e., similar users behave similarly \cite{Koren2009MF,Rendle2020}. Early matrix factorization (MF) based techniques linearly model the user-item interaction with inner product of user and item embeddings that are extracted from factor matrices \cite{Koren2009MF,rendle2010factorization}. To overcome the drawbacks of the MF based models that oversimplify the nonlinearity of user-item interactions, recently, various kinds of deep learning based models have been proposed to learn comprehensive representations for users and items, and capture the nonlinear user/item relationships \cite{zhang2019deep}. For example, Cheng et al. \cite{cheng2016wide} and He et al. \cite{He2017Neural} propose the Wide\&Deep model and the Neural Collaborative Filtering model, respectively, which can model the nonlinearity of user-item interactions with a multilayer perceptron. However, the traditional recommender systems often assume the user-item interaction data are noise-free, and lack the consideration on robustness of the models, which makes them likely fail to capture users' true preference from the data with perturbations \cite{Yuan2019Adversarial}.

\subsection{Robust Recommendation}

The existing works on robust recommendation roughly follow two lines, where one line is to improve the recommendation robustness by injecting noise to input or model parameters during model training, and the other line is to adopt a generative model like VAE to infer a latent representation space from which robust embeddings of users and items can be generated.

\subsection{Noise Injection Based methods for Robust Recommendation}
The existing methods for robust recommendation \cite{Yao2016Collaborative, He2018Adversarial, Yuan2019Adversarial, Li2020Adversarial} often inject extra noise to training data or model parameters to deal with noisy user-item interactions, which roughly fall into two classes. One class of methods, such as CDAE \cite{Yao2016Collaborative}, use Denoising Auto-encoder (DAE) \cite{Vincent2008Extracting} for generating robust embeddings of users and items, which adds random drop-out noise in user-item interaction vectors and trains an auto-encoder based on intentionally corrupted input with the objective of minimizing reconstruction errors. 

The other class of the noise based methods introduces adversarial noise as well as adversarial training to improve the model robustness \cite{goodfellow2014generative,He2018Adversarial,Tang2018Adversarial,Yuan2019Adversarial}. He et al. propose an Adversarial Personalized Ranking (APR) model which can enhance the pairwise ranking method BPR \cite{Rendle2012BPR} by performing adversarial training \cite{He2018Adversarial}. Tang et al. propose an Adversarial Multimedia Recommendation (AMR) model for robust recommendation of images, which is trained to defend an adversary of perturbations to the target image \cite{Tang2018Adversarial}. Yuan et al. propose a general adversarial training framework, which can improve both the robustness and the overall performance of NN-based recommendation models \cite{Yuan2019Adversarial}. 

There are two main defects in the above two classes of noise injection based methods. First, the model robustness depends on a fixed noise injection level set beforehand, which ignores the personalization of the noise reduction for different users. Second, for the adversarial noise based methods, it is hard to choose a proper adversarial noise level for the tradeoff between the overall performance and the robustness of the models, and an over strong adversarial noise level may impair the recommendation performance of the models.

\subsection{Variational Auto-encoder Based methods for Robust Recommendation}

Recently, due to the impressive power of VAE in representation learning in the fields of computer vision and network embedding \cite{Meng2019Co-Embedding, Bojchevski2018Deep, Santos2016Multilabel}, a line of VAE based methods have been proposed for robust collaborative filtering \cite{Bai2019Collaborative, He2018Collaborative, Li2017Collaborative, shenbin2020recvae, liang2018variational}. For example, He et al. propose an additional variational auto-encoder which can generate robust embeddings encoding side information of items, including content information and tag information \cite{He2018Collaborative}. Li et al. propose a collaborative variational auto-encoder (CVAE) for robust recommendation of multimedia, where VAE is used to generate latent representations for multimedia content \cite{Li2017Collaborative}. Shenbin et al. propose a Recommender VAE (RecVAE) model with a new architecture for the encoder network that can be trained with corrupted implicit user-item interaction vectors \cite{shenbin2020recvae}. However, as we have mentioned before, VAE based methods likely leads to less expressive models that are unable to handle the multi-modality of the distributions of user preference. At the same time, Makhzani et al. propose the Adversarial Auto-encoder (AAE) model \cite{Makhzani2016Adversarial} which can be used for variational inference. Similar to our model, AAE also uses an adversarial training to regularize the variational inference. However, different from our model where the posterior of each user is separately regularized (Equation (\ref{Eq_KL})), AAE regularizes the aggregated (averaged) posterior $q(\boldsymbol{x})$ to be close to the prior, i.e., minimizes 
\begin{equation}
\text{KL}\big(q(\boldsymbol{x})=\int_{\boldsymbol{u}}q(\boldsymbol{x}|\boldsymbol{u}) p(\boldsymbol{u}), p(\boldsymbol{x})\big),
\end{equation}
which deviates from the VAE's optimization objective of improving ELBO and limits its ability to capture multi-modality.

\section{CONCLUSION}

To overcome the defects of the existing methods for robust recommendation, we propose a novel Dual Adversarial Variational Embedding (DAVE) model which is able to provide the personalized noise reduction and capture the multi-modality of the preference distributions, by combining the advantages of VAE and adversarial training. Particularly, to provide the personalized noise reduction for different users and items, we introduce two VAEs, to infer a unique embedding distribution for each user and item, respectively. Due to the variational inference power of VAEs, the different noise levels of users and items can be adaptively captured by their own embedding distributions from which robust embeddings can be drawn. To improve the model expressiveness, we further introduce two GANs to DAVE. Due to the regularization offered by the adversarial training between the discriminators and the variational inference networks, DAVE is expressive enough to approximate the preference distributions with multi-modality. At last, the extensive experiments conducted on real datasets verify the effectiveness of DAVE on robust recommendations.

\section*{Acknowledgment}

This work is supported by National Natural Science Foundation of China under grant 61972270. This work is also supported in part by NSF under grants III-1763325, III-1909323, and SaTC-1930941.

\bibliographystyle{abbrv}
\bibliography{DAVE}

\begin{IEEEbiography}[{\includegraphics[width=1in,height=1.25in,clip,keepaspectratio]{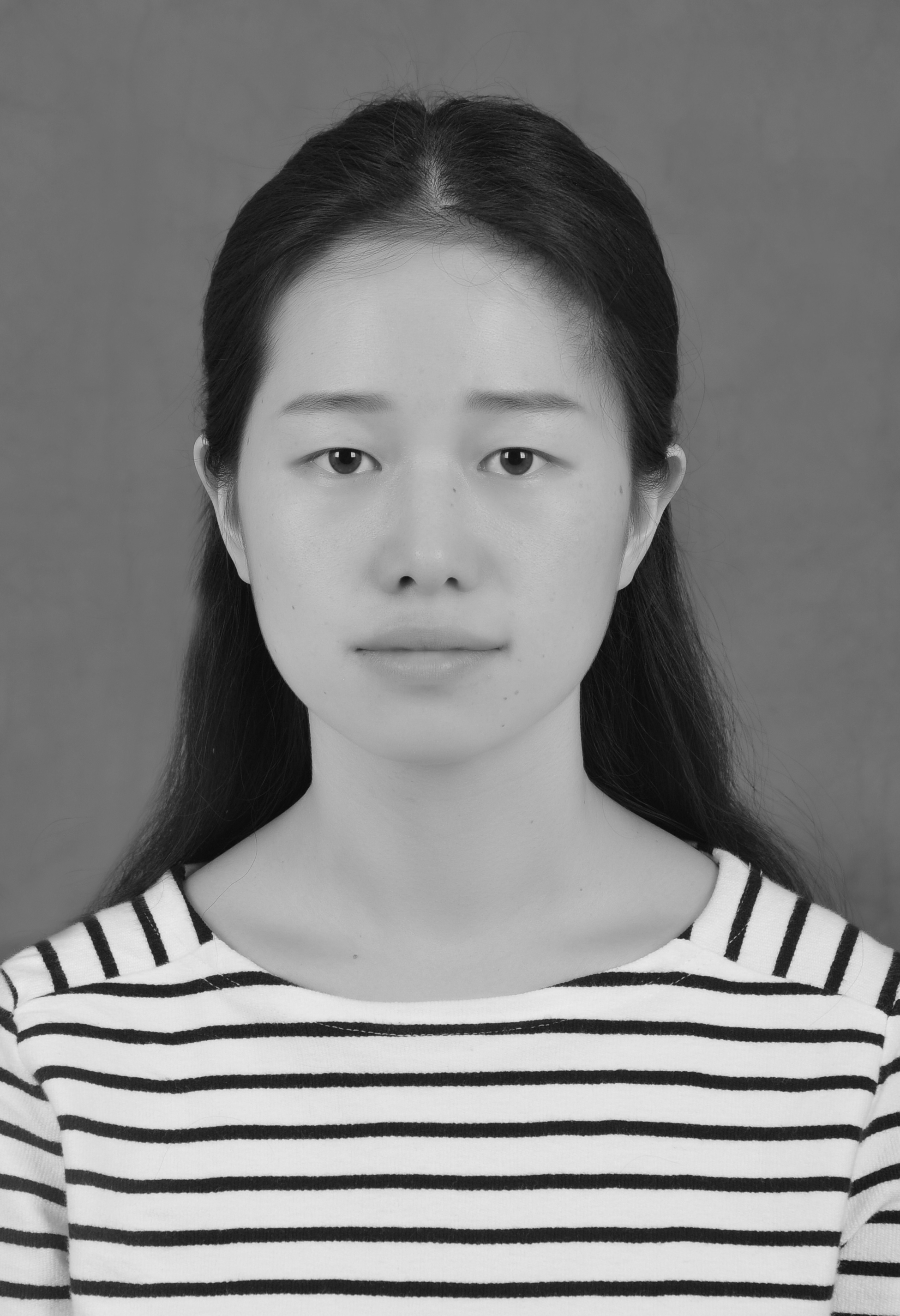}}]{Qiaomin Yi} obtained her bachelor's degree from the School of Information Engineering, Northwest A\&F University, China, in 2018. She is now pursuing the master's degree in the School of Computer Science, Sichuan University, China. Her research interests include data mining and recommender systems. 
\end{IEEEbiography}

\begin{IEEEbiography}[{\includegraphics[width=1in,height=1.25in,clip,keepaspectratio]{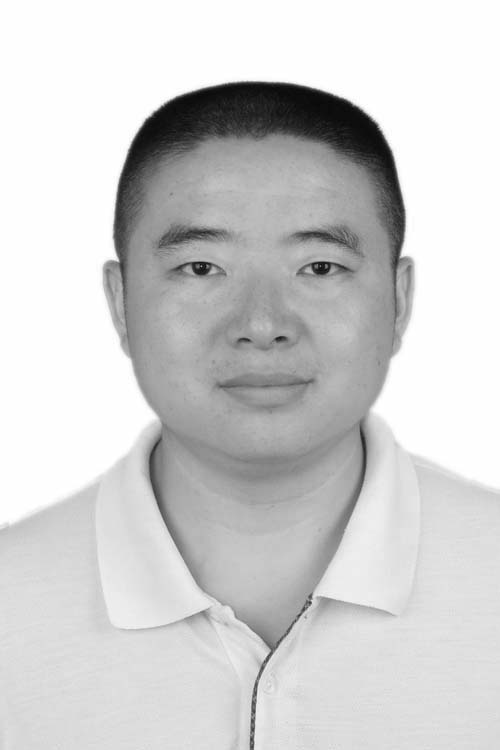}}]{Ning Yang} is an associate professor at Sichuan University, China. He obtained his Ph.D degree in Computer Science from Sichuan University in 2010. His research interests include recommender systems and social media mining.
\end{IEEEbiography}

\begin{IEEEbiography}[{\includegraphics[width=1in,height=1.25in,clip,keepaspectratio]{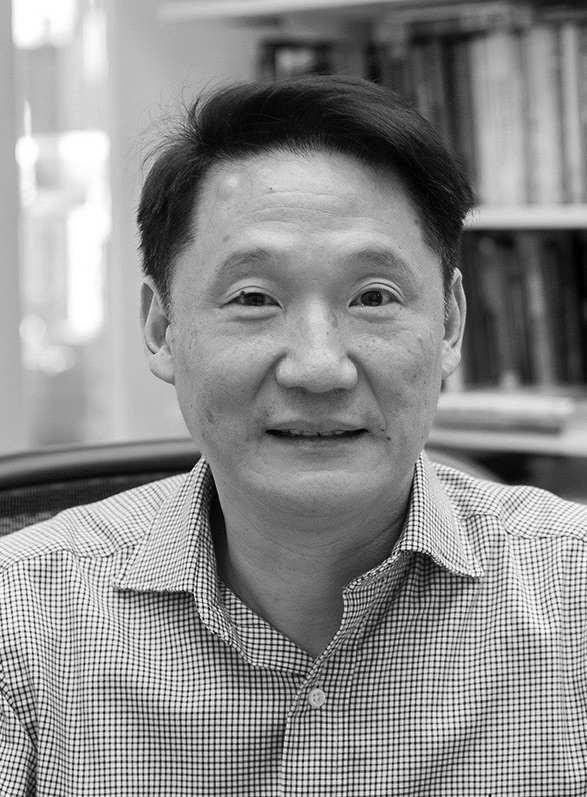}}]{Philip S. Yu} received the PhD degree in electrical engineering from Stanford University. He is a distinguished professor in computer science at the University of Illinois at Chicago and is also the Wexler chair in information technology. His research interests include big data, data mining, and social computing. He is a fellow of the ACM and the IEEE.
\end{IEEEbiography}
\vfill

\end{document}